\documentclass[pre,twocolumn,amsmath,amssymb,floatfix,superscriptaddress,nopacs,nofootinbib]{revtex4}

\usepackage{graphicx}
\usepackage{latexsym}
\usepackage{amsmath}
\usepackage{amssymb}
\usepackage{amsfonts}
\usepackage{bm}
\usepackage{color}

\newcommand{\Fcal}{\ensuremath{{\cal F}}}
\newcommand{\Hcal}{\ensuremath{{\cal H}}}
\newcommand{\ddiff}{\ensuremath{\mathrm{d}}}
\newcommand{\bfzero}{\ensuremath{{\bf 0}}}
\newcommand{\evec}{\mbox{$\bf e$}}
\newcommand{\qvec}{\mbox{$\bf q$}}
\newcommand{\rvec}{\mbox{$\bf r$}}

\newcommand{\uvec}{\mbox{$\bf u$}}
\newcommand{\vvec}{\mbox{$\bf v$}}
\newcommand{\Avec}{\mbox{$\bf A$}}
\newcommand{\Bvec}{\mbox{$\bf B$}}
\newcommand{\Cvec}{\mbox{$\bf C$}}
\newcommand{\Tvec}{\mbox{$\bf T$}}
\newcommand{\qhat}{\ensuremath{{\hat{q}}}}
\newcommand{\rhat}{\ensuremath{{\hat{r}}}}
\newcommand{\qhatvec}{\ensuremath{{\hat{\bf q}}}}
\newcommand{\rhatvec}{\ensuremath{{\hat{\bf r}}}}
\newcommand{\thetar}{\ensuremath{\theta_r}}
\newcommand{\thetaq}{\ensuremath{\theta_q}}
\newcommand{\Tgen}{\ast}
\newcommand{\la}{\left<}
\newcommand{\ra}{\right>}
\newcommand{\kB}{k_\mathrm{B}}
\newcommand{\kBT}{k_\mathrm{B}T}
\newcommand{\agrid}{a_{\mathrm{grid}}}
\newcommand{\nLgrid}{n_L}
\newcommand{\nVgrid}{n_V}
\newcommand{\tincr}{\delta \tau}
\newcommand{\tsamp}{\Delta \tau}

\newcommand{\Nt}{\ensuremath{N_\mathrm{t}}}
\newcommand{\Nc}{\ensuremath{N_\mathrm{c}}}

\newcommand{\sigabimp}{\delta\sigma_{\alpha\beta}}
\newcommand{\sigcdimp}{\delta\sigma_{\gamma\delta}}
\newcommand{\rref}{\tilde{\rvec}}
\newcommand{\ulongi}{u_{\mathrm{L}}}
\newcommand{\utrans}{u_{\mathrm{T}}}
\newcommand{\urmsq}{u_{\mathrm{rms}}}
\newcommand{\epsab}{\varepsilon_{\alpha\beta}}
\newcommand{\epsabhat}{\hat{\varepsilon}_{\alpha\beta}}
\newcommand{\epscd}{\varepsilon_{\gamma\delta}}
\newcommand{\epsxx}{\varepsilon_{11}}
\newcommand{\epsyy}{\varepsilon_{22}}
\newcommand{\epsxy}{\varepsilon_{12}}

\newcommand{\isfab}{\varepsilon^{\circ}_{\alpha\beta}}
\newcommand{\isfcd}{\varepsilon^{\circ}_{\gamma\delta}}
\newcommand{\isfxx}{\varepsilon^{\circ}_{11}}
\newcommand{\isfxy}{\varepsilon^{\circ}_{12}}
\newcommand{\isfyx}{\varepsilon^{\circ}_{21}}
\newcommand{\isfyy}{\varepsilon^{\circ}_{22}}
\newcommand{\isflongi}{\varepsilon_{\mathrm{L}}}
\newcommand{\isftrans}{\varepsilon_{\mathrm{T}}}
\newcommand{\cabcd}{c_{\alpha\beta\gamma\delta}}
\newcommand{\cxxxx}{c_{1111}}
\newcommand{\cyyyy}{c_{2222}}
\newcommand{\cxxyy}{c_{1122}}
\newcommand{\cxyxy}{c_{1212}}
\newcommand{\cxxxy}{c_{1112}}
\newcommand{\cxyyy}{c_{1222}}
\newcommand{\cone}{c_{\mathrm{L}}}
\newcommand{\ctwo}{c_{\mathrm{N}}}
\newcommand{\cthree}{c_{\perp}}
\newcommand{\cfour}{c_{\mathrm{T}}}
\newcommand{\cabcdprime}{c^{\prime}_{\alpha\beta\gamma\delta}}
\newcommand{\cxxxxprime}{c^{\prime}_{1111}}
\newcommand{\cyyyyprime}{c^{\prime}_{2222}}
\newcommand{\cxxyyprime}{c^{\prime}_{1122}}
\newcommand{\cxyxyprime}{c^{\prime}_{1212}}

\newcommand{\Tglass}{T_{\mathrm{g}}}
\newcommand{\Jone}{J_1}
\newcommand{\Jtwo}{J_2}
\newcommand{\scut}{s_\mathrm{cut}}
\newcommand{\smin}{s_\mathrm{min}}

\newcommand{\Eabcd}{\ensuremath{E_{\alpha\beta\gamma\delta}}}

\begin{document}

\title{Strain correlation functions in isotropic elastic bodies:\\
Large wavelength limit for two-dimensional systems}

\author{J.P.~Wittmer}
\email{joachim.wittmer@ics-cnrs.unistra.fr}
\affiliation{Institut Charles Sadron, Universit\'e de Strasbourg \& CNRS, 23 rue du Loess, 67034 Strasbourg Cedex, France}
\author{A.N. Semenov}
\affiliation{Institut Charles Sadron, Universit\'e de Strasbourg \& CNRS, 23 rue du Loess, 67034 Strasbourg Cedex, France}
\author{J. Baschnagel}
\affiliation{Institut Charles Sadron, Universit\'e de Strasbourg \& CNRS, 23 rue du Loess, 67034 Strasbourg Cedex, France}

\date{\today}

\begin{abstract}
Strain correlation functions in two-dimensional isotropic elastic bodies are shown both theoretically
(using the general structure of isotropic tensor fields) and 
numerically (using a glass-forming model system) 
to depend on the coordinates of the field variable 
(position vector $\rvec$ in real space or wavevector $\qvec$ in reciprocal space)
and thus on the direction of the field vector and the orientation of the coordinate system.
Since the fluctuations of the longitudinal and transverse components of the strain field 
in reciprocal space are known in the long-wavelength limit from the equipartition theorem, 
all components of the correlation function tensor field are imposed 
and no additional physical assumptions are needed.
An observed dependence on the field vector direction thus cannot be 
used as an indication for anisotropy or for a plastic rearrangement.
This dependence is different for the associated strain response field 
containing also information on the localized stress perturbation.
\end{abstract}
\maketitle

\section{Introduction}
\label{intro}

\begin{figure}[t]
\centerline{\resizebox{1.0\columnwidth}{!}{\includegraphics*{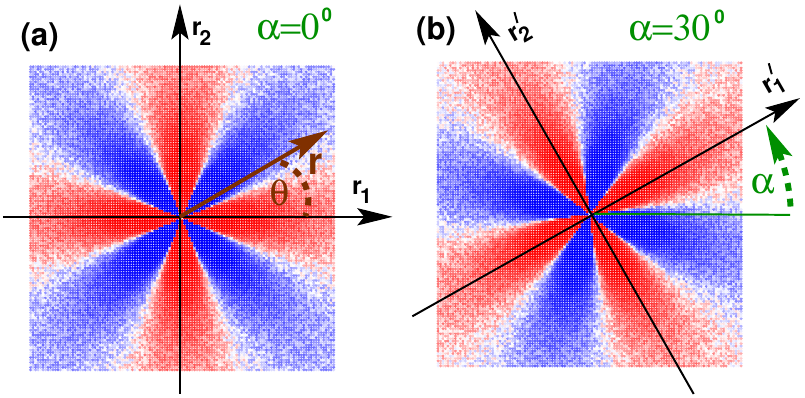}}}
\caption{Autocorrelation function $\cxyxy(\rvec)$ of the strain field component $\epsxy(\rvec)$ 
obtained from our colloidal glasses in two dimensions:
{\bf (a)} Unrotated frame with coordinates $(r_1,r_2)$,
{\bf (b)} frame ($r_1^{\prime},r_2^{\prime})$ rotated by an angle
$\alpha=30^{\circ}$ (rotations marked by ``$\prime$").
Albeit the system is isotropic, the correlation function is strongly
angle dependent, revealing an octupolar symmetry. 
While each pixel corresponds in {\bf (a)} and {\bf (b)} to the same spatial position $\rvec$,
the correlation functions differ by the angle $\alpha$.
$\cxyxy(\rvec)$ is positive (red) along the axes
and negative (blue) along the bisection lines of the respective axes.
}
\label{fig_intro}
\end{figure}

\subsection{General background}
\label{intro_background}

A tensor field assigns a tensor to each point of the mathematical space,
in our case for simplicity a two-dimensional Euclidean vector space with
Cartesian coordinates and an orthonormal tensor basis 
\cite{McConnell,Schouten,Schultz_Piszachich,spmP5a}.
Tensor fields are used in differential geometry \cite{McConnell}, general relativity \cite{Lambourne,Frankel_T}, 
in the analysis of stress and strain in materials \cite{ChaikinBook,LandauElasticity,TadmorCMTBook} 
and in numerous other applications in science and engineering. 
Tensor fields are experimentally \cite{Klix12,Klix15} or numerically 
\cite{Fuchs17,Fuchs18,Fuchs19,Lemaitre15,Lemaitre18,lyuda18,lyuda22a,MM22,spmP5a,Debregeas03,WXBB15,Weeks15,Bocquet04,Bocquet09,Lemaitre13,Barrat13,Barrat14c,FlennerSzamel15b,Fuchs16,Fuchs18b,Reichman21,Reichman21b} 
probed by means of correlation functions \cite{AllenTildesleyBook,HansenBook,ForsterBook} 
of their components and, importantly, these correlation functions are themselves components of tensor fields
\cite{spmP5a}. See Appendix~\ref{ten} for a brief review.
Assuming translational invariance, 
correlation functions are naturally best analyzed,
both for theoretical \cite{Fuchs17,Fuchs18,Fuchs19,lyuda18} and numerical 
\cite{AllenTildesleyBook,lyuda22a,spmP5a} reasons,
in a first step as functions of the wavevector $\qvec$ in reciprocal space.
The dependence on the spatial field vector $\rvec$ in real space can then be deduced 
(cf.~Appendix~\ref{app_FT}) in a second step by inverse Fourier transformation (FT). 
This was done, e.g., in our recent analysis \cite{spmP5a} of the spatial correlations 
of the (time-averaged) stress tensor fields in amorphous glasses formed by polydisperse Lennard-Jones (pLJ) 
particles deep in the glass regime (cf.~Sec.~\ref{tech_algo}).
It can thus be shown that all stress correlation functions
(both in reciprocal as in real space) can be described by means of 
{\em one} ``Invariant Correlation Function" (ICF) in reciprocal space characterizing 
the typical ensemble fluctuations of the quenched normal stress components 
in reciprocal space perpendicular to the wavevector $\qvec$.
Under additional but rather general assumptions \cite{spmP5a}
this ICF is given in the large-wavelength limit by a thermodynamic quantity, 
the equilibrium Young modulus of the system.

\subsection{Investigated case study}
\label{intro_casestudy}

As another example of the general procedure we shall investigate in the present work
the correlation functions $\cabcd(\rvec)=\Fcal^{-1}[\cabcd(\qvec)]$
of the instantaneous strain tensor field $\epsab(\rvec)$ in real space.
These may be readily obtained \cite{AllenTildesleyBook}
from the components of the tensor field 
\begin{equation}
\cabcd(\qvec) = \la \epsab(\qvec) \epscd(-\qvec) \ra
\label{eq_intro_cabcd_def}
\end{equation}
in reciprocal space with $\epsab(\qvec) = \Fcal[\epsab(\rvec)]$ 
being the Fourier transformed strain tensor field components. 
(The average $\la \ldots \ra$ will be specified below.)
An example for the autocorrelation function $\cxyxy(\rvec)$ of the shear strain $\epsxy(\rvec)$
is given in Fig.~\ref{fig_intro} 
for the same two-dimensional model system already used in Refs.~\cite{lyuda22a,spmP5a}.
Interestingly, the correlation function is seen to strongly depend both on the orientation of the field vector $\rvec$ 
(panel {\bf (a)}) and on the rotation angle $\alpha$ of the coordinate system (panel {\bf (b)}). 
Since the simulated system can be shown to be perfectly isotropic down to a few particle diameters 
\cite{WXP13,lyuda19a,spmP1,spmP2,spmP5a,lyuda22a}, these findings beg for an explanation.  
Expanding on our recent work on stress correlations \cite{lyuda18,lyuda22a,spmP5a}, 
this behavior can be traced back to the fact that correlation functions 
of tensor fields of isotropic systems must be components of a generic {\em isotropic} tensor field
(cf. Sec.~\ref{theo_cabcd}).
This field is shown below (cf.~Sec.~\ref{tech_corr})
to be completely described in terms of {\em two} ICFs
$\cone(q)$ and $\cfour(q)$ in reciprocal space ($q=|\qvec|$ being the magnitude of the wavevector).
These ICFs characterize the independent fluctuations of the longitudinal and 
transverse strain components $\isflongi(\qvec)$ and $\isftrans(\qvec)$.
Due to the equipartition theorem of statistical physics 
$\cone(q)$ and $\cfour(q)$ are given by \cite{ChaikinBook,Klix12,Klix15,WXBB15}
\begin{equation}
\beta V \cone(q) \to \frac{1}{\lambda + 2 \mu} \mbox{ and } \beta V \cfour(q) \to \frac{1}{4\mu} 
\mbox{ for } q \to 0
\label{eq_intro_ICF_qlow}
\end{equation}
in the large-wavelength limit with 
$\beta = 1/\kBT$ being the inverse temperature, $V$ the $d$-dimensional volume of the system
and $\lambda$ and $\mu$ two macroscopic Lam\'e coefficients \cite{ChaikinBook,LandauElasticity}.
All strain correlation functions are thus imposed on large scales.
In turn this explains without any additional physical input
the octupolar pattern\footnote{See, e.g., the wikipedia entries on quadrupoles and
            general multipolar expansions as used, say, in electrostatics.
            For the planar harmonic basis functions $\cos(p \theta)$ or $\sin(p \theta)$
            a monopole corresponds to $p=0$,
            a dipole to $p=1$, a quadrupole to $p=2$ and an octupole to $p=4$.}
observed in Fig.~\ref{fig_intro} (cf. Sec.~\ref{res_Cabcd_r} and Appendix~\ref{app_Cabcd})
and shows that strain correlations in elastic bodies must necessarily be long-ranged.
This is different for the closely related but distinct tensorial response field
being the tensorial product of correlation functions and the imposed
tensorial perturbation. As emphasized in Sec.~\ref{theo_source} and Sec.~\ref{response},
the response field thus contains additional information due to the source term and its symmetry.

\subsection{Outline}
\label{intro_outline}

We begin in Sec.~\ref{theo} with some general theoretical considerations on isotropic tensor fields. 
Technical points concerning the model system and the data production of tensorial fields
on discrete grids are discussed in Sec.~\ref{tech}.
This is followed in Sec.~\ref{res} by the presentation of our main numerical results.
The strain response due to an imposed stress point source is discussed in Sec.~\ref{response}.
A summary and an outlook are given in Sec.~\ref{conc}.
More details may be found in the Appendix both 
on the theoretical background 
(cf.~Appendices~\ref{ten} and \ref{app_Cabcd})
and on computational issues 
(cf.~Appendices~\ref{app_FT} and \ref{app_comp}). 

\section{General considerations}
\label{theo}

\subsection{Isotropic tensors and tensor fields}
\label{theo_isofields}

Isotropic systems, such as generic isotropic elastic bodies \cite{ChaikinBook,LandauElasticity,TadmorCMTBook},
simple and complex fluids \cite{FerryBook,HansenBook,RubinsteinBook,DoiEdwardsBook},
amorphous metals and glasses 
\cite{DonthGlasses,Bocquet04,Bocquet09,Lemaitre13,Barrat14c,FlennerSzamel15b,Fuchs16,Fuchs18b,Reichman21,Reichman21b}, 
polymer networks and gels \cite{FerryBook,RubinsteinBook}, 
foams and emulsions \cite{Debregeas03,Weeks15}
or, as a matter of fact, our entire universe \cite{Lambourne} are described at least on some scales
by {\em isotropic tensors} and {\em isotropic tensor fields} (cf.~Appendix~\ref{ten_iso}) 
\cite{McConnell,TadmorCMTBook,Schultz_Piszachich}. 
It is well known \cite{Schultz_Piszachich,TadmorCMTBook} that the components of isotropic tensors 
remain unchanged under an orthogonal coordinate transformation (including rotations and reflections).
For instance, 
\begin{equation}
\Eabcd^{\Tgen} = \Eabcd
\label{eq_theo_tensor_criterion}
\end{equation}
for the forth-order elastic modulus tensor of an isotropic body (cf.~Appendix~\ref{app_comp_elastmacro}) 
\cite{LandauElasticity,TadmorCMTBook} with ``$\Tgen$" marking an arbitrary orthogonal transformation 
(cf.~Appendix~\ref{ten_intro}).
This implies (cf.~Appendix~\ref{ten_field}) that $\Eabcd$ is given by two invariants,
e.g., the two Lam\'e coefficients $\lambda$ and $\mu$. 
Importantly, this does {\em not} hold for isotropic tensor fields \cite{Schultz_Piszachich,lyuda18,spmP5a}. 
For instance, for a forth-order correlation function in reciprocal space the isotropy condition becomes 
\begin{equation}
\cabcd^{\Tgen}(\qvec) = \cabcd(\qvec^{\Tgen})
\label{eq_theo_field_criterion}
\end{equation}
with $\qvec^{\Tgen}$ being the ``actively" transformed wavevector (cf.~Appendix~\ref{ten_iso}).

\subsection{Structure of isotropic correlation functions}
\label{theo_cabcd}

Assuming in addition the system to be achiral and two-dimensional (cf. Appendix~\ref{ten_sym})
it can be shown \cite{spmP5a} that correlation functions of second-order tensor field components must take 
the following mathematical structure 
\begin{eqnarray}
\cabcd(\qvec) & = & i_1(q) \ \delta_{\alpha\beta} \delta_{\gamma\delta}  \label{eq_theo_cabcd} \\
& + & i_2(q) \left[
\delta_{\alpha\gamma} \delta_{\beta\delta} + \delta_{\alpha\delta} \delta_{\beta\gamma}
\right] \nonumber \\
& + & i_3(q) \left[
\qhat_{\alpha} \qhat_{\beta}\delta_{\gamma\delta} + \qhat_{\gamma} \qhat_{\delta}\delta_{\alpha\beta} 
\right] \nonumber \\
& + &  i_4(q) \ \qhat_{\alpha} \qhat_{\beta} \qhat_{\gamma} \qhat_{\delta} \nonumber
\end{eqnarray}
in terms of four ICFs $i_n(q)$, the coordinates $\qhat_{\alpha}$ of the normalized wavevector $\qhatvec$
and the Kronecker symbol $\delta_{\alpha\beta}$.
Legitimate correlation functions of isotropic systems may thus depend on $\qhat_{\alpha}$ 
and, hence, on the orientation of the wavevector and of the coordinate system.
While the isotropy of the system may not be manifested by {\em one} correlation function,
it is crucial for the structure of the {\em complete set} of {\em all} 
correlation functions given by Eq.~(\ref{eq_theo_cabcd}).
We note finally that it is useful to express the above ICFs in terms of an alternative set of ICFs 
$\cone(q)$, $\cfour(q)$, $\cthree(q)$ and $\ctwo(q)$ given by
\begin{eqnarray}
i_1(q) & = & \ctwo(q) - 2\cfour(q) \label{eq_theo_in2cn} \\
i_2(q) & = & \cfour(q) \nonumber \\
i_3(q) & = & \cthree(q) - \ctwo(q)+ 2\cfour(q)  \nonumber \\
i_4(q) & = & \cone(q) + \ctwo(q) - 2\cthree(q) - 4\cfour(q). \nonumber
\end{eqnarray}
See Appendix~\ref{ten_field} for more details.

\subsection{Planar harmonic basis functions}
\label{theo_harmonics}

Instead of using the components $\qhat_{\alpha}$ 
one may, quite generally, express all isotropic tensor fields 
in two dimensions in terms of the orthogonal planar harmonic basis functions 
$\cos(p \theta)$ and $\sin(p \theta)$
with $\qhat_1=\cos(\theta)$ and $\qhat_2=\sin(\theta)$ and $p=0$, $2$ and $4$. 
(See Appendix~\ref{app_Cabcd} for more details.)
For instance, it follows from Eq.~(\ref{eq_theo_cabcd}) that 
\begin{equation}
c_{1212}(\qvec) = i_2(q)+\frac{i_4(q)}{8} - \frac{i_4(q)}{8} \cos(4 \theta).
\label{eq_theo_c1212}
\end{equation}
Hence, if the invariant $i_4(q)$ is sufficiently large, 
$c_{1212}(\qvec)$ must reveal an octupolar pattern. 
Due to Eq.~(\ref{eq_q2r_fr_p}) derived in Appendix~\ref{app_FT_inverse},
this alternative representation is especially useful for performing the inverse FT to real space.
This also shows that the corresponding correlation function 
$c_{\alpha\beta\gamma\delta}(\rvec) = \Fcal^{-1}[c_{\alpha\beta\gamma\delta}(\qvec)]$ 
in real space must have the same mathematical properties.

\subsection{Response to point source}
\label{theo_response}

Let us consider the second order tensor field 
$R_{\alpha\beta}(\qvec)$ obtained by the contraction 
\begin{equation}
R_{\alpha\beta}(\qvec) = \frac{1}{V} \ \cabcd(\qvec) s_{\gamma\delta}
\label{eq_theo_res_A}
\end{equation}
with a symmetric but not necessarily isotropic tensor $s_{\alpha\beta}$
using the standard summation convention over repeated indices \cite{McConnell,Schultz_Piszachich}.
(For convenience we have introduced the system volume $V$.)
We shall call $R_{\alpha\beta}(\qvec)$ the ``response field" (in reciprocal space) 
and $s_{\alpha\beta}$ the ``point source tensor".
In fact, using Eq.~(\ref{eq_FT_B}) and Eq.~(\ref{eq_FT_convolution_q}) 
it is seen that in real space the tensor $s_{\alpha\beta}/V$
corresponds to a ``point source" $s_{\alpha\beta} \delta(\rvec)$
(using Dirac's delta function) and $R_{\alpha\beta}(\qvec)$ becomes
\begin{equation}
R_{\alpha\beta}(\rvec) = \Fcal^{-1}[R_{\alpha\beta}(\qvec)]=\cabcd(\rvec) s_{\gamma\delta} 
\label{eq_theo_res_AA}
\end{equation}
using $\cabcd(\rvec) = \Fcal^{-1}[\cabcd(\qvec)]$.
We shall say more about the specific linear strain response in real space in Sec.~\ref{response}
but focus here on the generic response in reciprocal space.
Being symmetric the source tensor may be diagonalized by a rotation of the coordinate system 
where $s_{12}=s_{21}=0$ and $s_{11}$ and $s_{22}$ become the two (in general not identical) eigenvalues. 
Hence,
\begin{equation}
R_{\alpha\beta}(\qvec) = \frac{1}{V} 
\left[ s_{11} c_{\alpha\beta11}(\qvec) + s_{22} c_{\alpha\beta22}(\qvec) \right].
\label{eq_theo_res_B}
\end{equation}
We emphasize that the sum must be taken over {\em all} eigenvalues of the source tensor,
i.e. two for the presented two-dimensional case. 
(The failure to sum properly over {\em all} tensorial contributions
to $R_{\alpha\beta}(\qvec)$ leads to incorrect angular dependences.)
Importantly, $R_{\alpha\beta}(\qvec)$ thus contains information over {\em both} the system,
characterized by the correlation functions, {\em and} the imposed source term.

\subsection{Different types of source terms}
\label{theo_source}

If we now assume that not only $\cabcd(\qvec)$ is an isotropic tensor field but
that, moreover, $s_{\alpha\beta}$ is isotropic, i.e. $s_{11}=s_{22}$,
the product theorem Eq.~(\ref{eq_ten_iso_AB2C}) discussed in Appendix~\ref{ten_iso}
implies that $R_{\alpha\beta}(\qvec)$ must also be an isotropic tensor field.
According to Eq.~(\ref{eq_ten_field_o2}) it is given by
\begin{equation}
R_{\alpha\beta}(\qvec) = k_1(q) \delta_{\alpha\beta} + k_2(q) \qhat_{\alpha} \qhat_{\beta}
\label{eq_theo_res_D}
\end{equation}
in terms of two invariants $k_1(q)$ and $k_2(q)$ which 
can in turn be expressed in terms of the invariants of
$c_{\alpha\beta\gamma\delta}(\qvec)$ and $s_{\alpha\beta}$.
$R_{\alpha\beta}(\qvec)$ can thus at most be quadrupolar ($p=2$). 
Specifically, 
\begin{equation}
R_{12}(\qvec) = k_2(q) \qhat_1 \qhat_2 \propto \sin(2\theta)
\label{eq_theo_res_isosource}
\end{equation}
which is distinct from $c_{1212}(\qvec)$, cf. Eq.~(\ref{eq_theo_c1212}).
Importantly, in many physical situations the source is in fact {\em not} isotropic
and thus in turn the response field {\em not} consistent with Eq.~(\ref{eq_theo_res_D}).
We remind that
according to a popular model of localized plastic failure
by means of ``shear transformation zones" \cite{Argon79,Bocquet04,Tanguy11,Barrat18,Langer98}
two orthogonal twin force dipoles of {\em opposite} signs may be imposed at the 
origin.\footnote{Let 
us impose in a rotated coordinate system at $\alpha=-\pi/4$
a symmetric source tensor with a finite ``shear" $s^{\prime}_{12}=s$ and vanishing
diagonal components $s_{11}^{\prime} = s_{22}^{\prime} =0$.
Using Eq.~(\ref{eq_ten_intro_orthtrans}) this implies
$s_{11}=-s_{22}=s$ and $s_{12}=0$ in the original coordinate system at $\alpha=0$.}
This suggests to consider the case $s_{11}=-s_{22}$.
It follows then from Eq.~(\ref{eq_theo_cabcd}) and Eq.~(\ref{eq_theo_res_B}) that
\begin{equation}
R_{12}(\qvec) \propto i_4(q) \ \qhat_1 \qhat_2 (\qhat_1^2-\qhat_2^2) \propto \sin(4\theta).
\label{eq_theo_res_asosource}
\end{equation}
The (non-isotropic) response field $R_{12}(\qvec)$ thus is in this case octupolar 
as the correlation field $c_{1212}(\qvec)$, however, shifted by an angle $\pi/8$.
It is readily seen by inverse FT that the same general behavior applies in real space.

\section{Technical issues}
\label{tech}

\subsection{Algorithm, configurations and frames}
\label{tech_algo}

We investigate amorphous glasses in two dimensions formed by pLJ particles 
\cite{WXP13,lyuda19a,spmP1,spmP2,spmP5a,lyuda21b,lyuda22a} 
which are sampled by means of Monte Carlo (MC) simulations \cite{AllenTildesleyBook}.
See Appendix~\ref{app_comp} for details (Hamiltonian, units, cooling and equilibration procedure, 
data production, generation and analysis of tensor fields on discrete grids). 
We focus on systems containing $n=10000$ and $40000$ particles at a working temperature $T=0.2$.
This is much lower than the glass transition temperature $\Tglass \approx 0.26$ \cite{WXP13},
i.e. for any computationally feasible production time the systems behave 
as solid elastic bodies \cite{spmP1}.
$\Nc=200$ completely independent configurations $c$ are prepared using a mix of local and swap MC hopping 
moves \cite{Berthier17,spmP1} while the presented data are computed using local MC moves only.
For each $c$ we store time-series containing $\Nt=10000$ ``frames" $t$ computed using equidistant time intervals.
As described in Appendix~\ref{app_comp_elastmacro}, the elastic modulus tensor $\Eabcd$ is isotropic 
and determined by the two Lam\'e coefficients $\lambda \approx 38$ and $\mu \approx 14$ \cite{WXP13,spmP1}.

\begin{figure}[t]
\centerline{\resizebox{.80\columnwidth}{!}{\includegraphics*{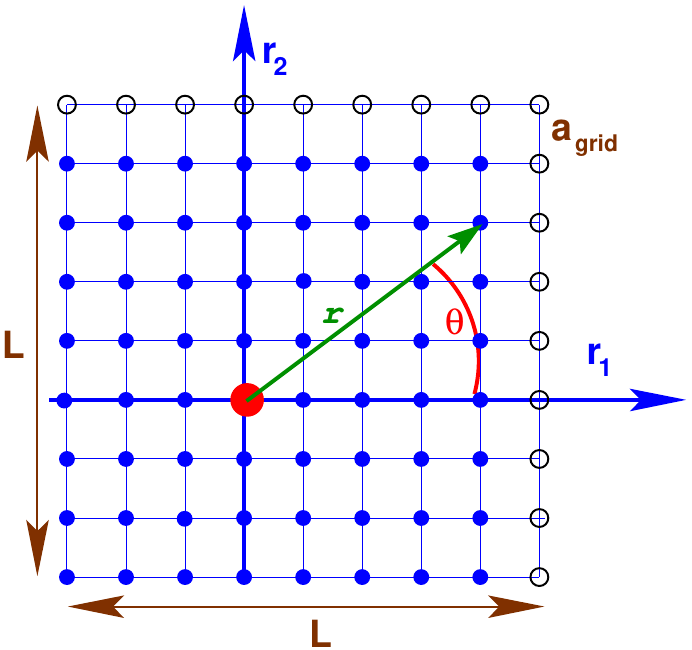}}}
\caption{Two-dimensional ($d=2$) square lattice with $\agrid$ being the lattice constant 
and $\nLgrid=L/\agrid$ the number of grid points in one spatial dimension. 
The filled circles indicate microcells of the principal box,
the open circles some periodic images.
The spatial position $\rvec$ of a microcell is either given by the $r_1$- and 
$r_2$-coordinates (in the principal box) or by the distance $r=|\rvec|$ 
from the origin (large circle) and the angle $\theta$.
}
\label{fig_grid}
\end{figure}

\subsection{Sampled discrete tensorial fields}
\label{tech_fields}

As shown in Fig.~\ref{fig_grid}, a discrete square grid is used
to store and to manipulate the various fields needed for the microscopic description.
The standard lattice constant for the grid in real space is $\agrid \approx 0.2$.
The displacement field $\uvec(\rvec)$ in real space is determined for each frame $t$
using a standard method \cite{Klix12,Klix15,WXBB15}
from the displacement vector of each particle
using as reference position the time-averaged particle position (cf.~Appendix~\ref{app_comp_uq}).
We obtain then from the Fourier transformed displacement field
$\uvec(\qvec) = \Fcal[\uvec(\rvec)]$ the strain tensor field
\cite{LandauElasticity,lyuda18}
\begin{equation}
\epsab(\qvec) = \frac{i}{2} \left(q_{\beta} u_{\alpha}(\qvec) + q_{\alpha} u_{\beta}(\qvec)\right)
\label{eq_algo_epsab}
\end{equation}
in reciprocal space. Using the correlation function theorem for FTs (cf.~Appendix~\ref{app_FT})
the strain correlations functions in reciprocal space are given by Eq.~(\ref{eq_intro_cabcd_def})
where the average is taken over all $t$ and $c$.
Both strain and correlation function fields in reciprocal space are defined to be dimensionless
(cf. Appendix~\ref{app_FT_continuous}).
We emphasize by a  prime ``$\prime$" all tensor field components obtained in a coordinate system
rotated by an angle $\alpha$ (with $\alpha=0$ being the original unrotated system).
Specifically, the correlation functions
$\cabcdprime(\qvec) = \langle \epsab^{\prime}(\qvec) \epscd^{\prime}(-\qvec)\rangle$
are obtained using the components $u^{\prime}_{\alpha}(\rvec)$ and $q^{\prime}_{\alpha}$ 
in the rotated frame.

\subsection{Natural Rotated Coordinates}
\label{tech_NRC}

All the tensorial fields introduced above depend on the orientation of the coordinate system.
Importantly, we  consider these properties in a first step in ``Natural Rotated Coordinates" (NRC)
where for {\em each} wavevector $\qvec$ the coordinate system is rotated until the $1$-axis 
coincides with the $\qvec$-direction. We mark these new tensor field components by ``$\circ$" 
to distinguish them from standard rotated tensor field components (marked by primes ``$\prime$").
Note that $q_{\alpha}^{\circ} = q \delta_{1\alpha}$ for all wavevectors $\qvec$.
Using the components $q_{\alpha}^{\circ}$ and $u_{\alpha}^{\circ}(\qvec)$ 
we obtain (as before) the strain tensor $\epsab^{\circ}(\qvec)$.
Importantly, 
\begin{equation}
q_2^{\circ}=0 \ \Rightarrow \ \isfyy(\qvec)=0
\label{eq_tech_isfyy}
\end{equation}
in agreement with Eq.~(\ref{eq_algo_epsab}).
We thus only have two independent components of the strain tensor field in NRC.
We alternatively write for convenience  
$\ulongi(\qvec) \equiv u_1^{\circ}(\qvec)$, $\utrans(\qvec) \equiv u_2^{\circ}(\qvec)$,
$\isflongi(\qvec) \equiv \isfxx(\qvec)$ and $\isftrans(\qvec) \equiv \isfxy(\qvec) \equiv \isfyx(\qvec)$
for the longitudinal and transverse components of the displacement and strain tensor fields.
Note that 
\begin{eqnarray}
\isfxx(\qvec) & \equiv & \isflongi(\qvec) = i q \ulongi(\qvec) \mbox{ and }
\label{eq_tech_isflongi} \\
\isfxy(\qvec) \equiv \isfyx(\qvec) & \equiv & \isftrans(\qvec) = i q \utrans(\qvec)/2,
\label{eq_tech_isftrans} 
\end{eqnarray}
i.e. displacement and strain fields in NRC contain essentially the same information.

\subsection{Correlation functions in NRC}
\label{tech_corr}

The correlation functions $\cabcd^{\circ}(\qvec) \equiv \langle \isfab(\qvec) \isfcd(-\qvec)\rangle$ 
may for {\em finite} $\Nc$ not only depend on $q$ but also on $\qhatvec$. 
Consistently with Eq.~(\ref{eq_ten_Tabcd_d2_A}) and Refs.~\cite{lyuda18,lyuda22a,spmP5a} 
we thus operationally define the ICFs 
$\cone(q)  \equiv \langle \cxxxx^{\circ}(\qvec) \rangle_{\qhatvec}$,
$\cfour(q) \equiv \langle \cxyxy^{\circ}(\qvec) \rangle_{\qhatvec}$,
$\ctwo(q)  \equiv \langle \cyyyy^{\circ}(\qvec) \rangle_{\qhatvec}$ and
$\cthree(q)\equiv \langle \cxxyy^{\circ}(\qvec) \rangle_{\qhatvec}$
by averaging over all wavevectors with $|\qvec| \approx q$. 
However, $\isfyy(\qvec)=0$ implies immediately that 
\begin{equation}
\cyyyy^{\circ}(\qvec)=\cxxyy^{\circ}(\qvec)=\ctwo(q)=\cthree(q)=0 \mbox{ for } \forall \ \qvec.
\label{eq_ICFs_vanishing}
\end{equation}
The two remaining non-trivial ICFs $\cone(q)$ and $\cfour(q)$ are called, respectively, 
the ``longitudinal ICF" and the ``transverse ICF".
As already noted in the Introduction, according to the equipartition theorem 
$\cone(q)$ and $\cfour(q)$ are given for sufficiently large wavelengths by the Lam\'e coefficients 
$\lambda$ and $\mu$. The stated Eq.~(\ref{eq_intro_ICF_qlow}) can be readily obtained from published
work \cite{ChaikinBook,Klix12,Klix15,WXBB15} 
using Eq.~(\ref{eq_tech_isflongi}) and Eq.~(\ref{eq_tech_isftrans}) 
to substitute the displacement fields $\ulongi(\qvec)$ and $\utrans(\qvec)$ in NRC 
by the corresponding strain fields $\isflongi(\qvec)$ and $\isftrans(\qvec)$.

\section{Main numerical results}
\label{res}

\begin{figure}[t]
\centerline{\resizebox{0.9\columnwidth}{!}{\includegraphics*{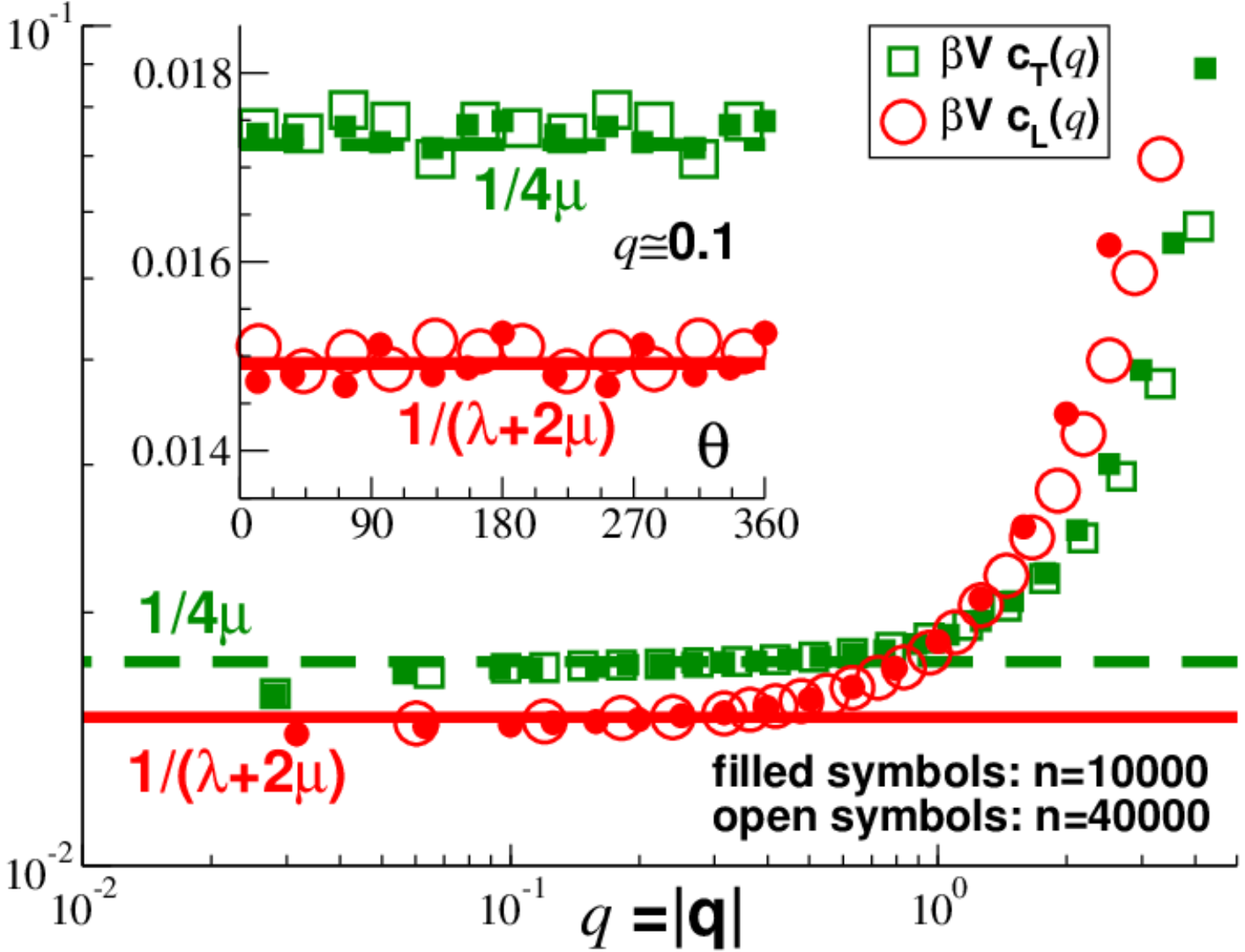}}}
\caption{Rescaled correlation functions in NRC and reciprocal space.
The bold dashed and solid lines indicate the expected low-$q$ limit Eq.~(\ref{eq_intro_ICF_qlow}).
Inset:
$\beta V \cxyxy^{\circ}(\qvec)$ and $\beta V \cxxxx^{\circ}(\qvec)$ 
{\em vs.} $\theta$ for $q \approx 0.1$.
Main panel:
Semi-logarithmic representation of ICFs $\beta V \cfour(q)$ and $\beta V \cone(q)$ {\em vs.} $q$.
}
\label{fig_Crot}
\end{figure}

\subsection{Measured longitudinal and transverse ICFs}
\label{res_ICFs}

We turn now to the numerical results of this work.
Figure~\ref{fig_Crot} focuses on the two non-vanishing correlation functions obtained in reciprocal space and NRC. 
All correlation functions are rescaled by $\beta V$ having thus the dimension of an inverse modulus.
As can be seen for the two indicated particle numbers $n$, a data collapse for different system sizes is observed, 
confirming the expected volume scaling.
The inset presents the (not yet spherically averaged) correlation functions
$\beta V \cxyxy^{\circ}(\qvec)$ and $\beta V \cxxxx^{\circ}(\qvec)$ as functions of 
the wavevector angle $\theta$ for one small wavevector with $q \approx 0.1$. 
As expected for isotropic systems, these correlation functions are $\theta$-independent 
(apart from a small noise contribution due to the finite number $\Nc$ of independent configurations).
The main panel presents the $\qhatvec$-averaged longitudinal and transverse ICFs 
$\beta V \cone(q)$ and $\beta V \cfour(q)$ as functions of $q$.
The expected large-wavelength limit Eq.~(\ref{eq_intro_ICF_qlow}) is indicated 
in both panels by bold horizontal lines. As can be seen in the main panel, 
it is well confirmed for $q \ll 1$ over at least one order of magnitude
where we have used the known values of $\lambda$ and $\mu$.
We remind that Eq.~(\ref{eq_intro_ICF_qlow}) has been used in various experimental and numerical studies 
\cite{Klix12,Klix15,WXBB15} to fit $\lambda$ and $\mu$.
$\cone(q)$ and $\cfour(q)$ characterize the typical 
length of the complex random variables $\isflongi(\qvec)$ and $\isftrans(\qvec)$.
Their distributions and correlations will be discussed elsewhere \cite{spmP5c}.
We note finally that the increase of the ICFs from the low-$q$ asymptotics visible for $q >1$ 
correlates with the deviation of the total static structure factor $S(q)$ from its
low-$q$ plateau (cf.~Fig.~\ref{fig_comp_uq}). 

\subsection{Correlation functions in reciprocal space}
\label{res_Cabcd_q}

While remaining in reciprocal space we consider next coordinate frames
which are either unrotated ($\alpha=0$) or rotated as in Fig.~\ref{fig_intro}(b) 
using the {\em same} angle $\alpha$ for all $\qvec$. 
According to Eq.~(\ref{eq_theo_cabcd}) the correlation functions $\cabcd(\qvec)$
of isotropic achiral systems in two dimensions depend quite generally on the four ICFs
$\cone(q)$, $\cfour(q)$, $\ctwo(q)$ and $\cthree(q)$.
Due to Eq.~(\ref{eq_ICFs_vanishing}) the last two of these ICFs must vanish 
while $\cone(q)$ and $\cfour(q)$ are given by Eq.~(\ref{eq_intro_ICF_qlow}). 
Let us introduce for later convenience the two ``creep compliances"
\begin{equation}
\Jone \equiv \frac{1}{\mu}-\frac{1}{\lambda+2\mu} \mbox{ and } \Jtwo \equiv \frac{2}{\lambda+2\mu}.
\label{eq_res_JoneJtwo}
\end{equation}
This yields in the original coordinates 
\begin{eqnarray}
\hspace*{-0.5cm}\beta V \cxyxy(\qvec) & \to  & \frac{\Jone}{8} \cos(4\theta) +\ldots \label{eq_res_cxyxy_q}\\
\beta V \cxxyy(\qvec) & \to & \frac{\Jone}{8} \cos(4\theta) + \ldots \label{eq_res_cxxyy_q} \\
-\frac{\beta V}{2} (\cxxxx(\qvec)+\cyyyy(\qvec)) & \to & \frac{\Jone}{8} \cos(4\theta) + \ldots 
\label{eq_res_cmean_q} \\
\frac{\beta V}{2} (\cxxxx(\qvec)-\cyyyy(\qvec)) & \to & \frac{\Jtwo}{4} \cos(2\theta)  
\label{eq_res_cdiff_q} 
\end{eqnarray}
for $q \to 0$. 
The dots mark irrelevant constant contributions.\footnote{The omitted constant terms correspond 
to localized $\delta(\rvec)$ contributions to strain correlation functions in real space. 
For example, such a contribution to $c_{1212}(\rvec)$ is $\frac{\Jone+\Jtwo}{8\beta} \delta(\rvec)$.}
%
See Appendix~\ref{app_Cabcd} for more details. 
For correlation functions $\cabcdprime(\qvec)$ in rotated coordinate systems 
one merely needs to substitute $\theta$ by $x=\theta-\alpha$.
\begin{figure}[t]
\centerline{\resizebox{0.9\columnwidth}{!}{\includegraphics*{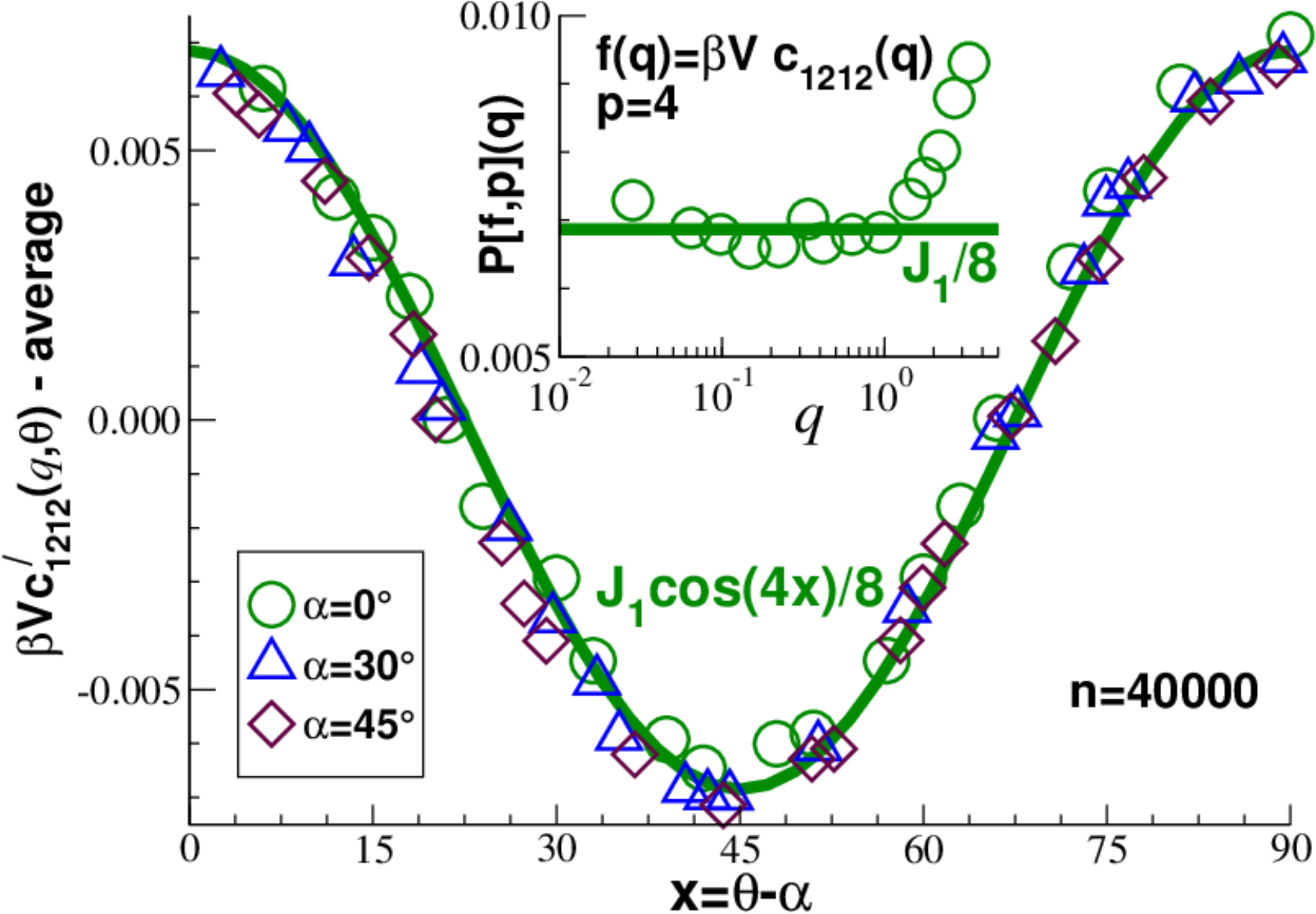}}}
\caption{Rescaled correlation function $f(\qvec) = \beta V \cxyxyprime(\qvec)$ for $n=40000$.
Main panel: Angle dependence of vertically shifted $f(\qvec)$ for $q \approx 0.1$.
Data collapse is observed using the reduced angle $x=\theta-\alpha$.
The bold solid line indicates the prediction, Eq.~(\ref{eq_res_cxyxy_q}).
Inset:
Comparison of $P[f,p](q)$ for $p=4$ with the predicted low-$q$ limit $\Jone/8$
(bold solid line).
}
\label{fig_Cabcd_q}
\end{figure}
These relations are put to a test in Fig.~\ref{fig_Cabcd_q} where we focus for clarity on the 
reduced shear-strain autocorrelation function $f(\qvec) =\beta V \cxyxyprime(\qvec)$ for $n=40000$.
The angular dependences are presented in the main panel for one wavevector in the low-$q$ limit.
Focusing on the first term in Eq.~(\ref{eq_res_cxyxy_q})
we have taken off the mean constant average over all $\theta$
(corresponding to the dots).
Importantly, all data for different $\alpha$ are seen to
collapse when plotted as a function of the scaling variable $x$. 
Obviously, this simple scaling (without characteristic angle) would not hold for anisotropic systems.
To obtain a precise test of the $q$-dependence of $\cabcd(\qvec)$
we project out the angular dependences using 
\begin{equation}
P[f,p](q) \equiv
2 \times \frac{1}{2\pi} \int_0^{2\pi} \ddiff \theta \ f(q,\theta) \cos(p \theta)
\label{eq_res_Pfp_def}
\end{equation}
for $p=2$ and $p=4$. For convenience the prefactor of the integral is chosen such that
$P[\cos(2\theta),2]=P[\cos(4\theta),4]= 1$.
The result for the shear-stress autocorrelation function with $p=4$ is shown in the inset of Fig.~\ref{fig_Cabcd_q}.
In agreement with Eq.~(\ref{eq_res_cxyxy_q}) the presented data
is given by $\Jone/8$ (solid line) for sufficiently small wavevectors.
Equivalent results have been obtained for the other correlation functions mentioned above.

\begin{figure}[t]
\centerline{\resizebox{0.9\columnwidth}{!}{\includegraphics*{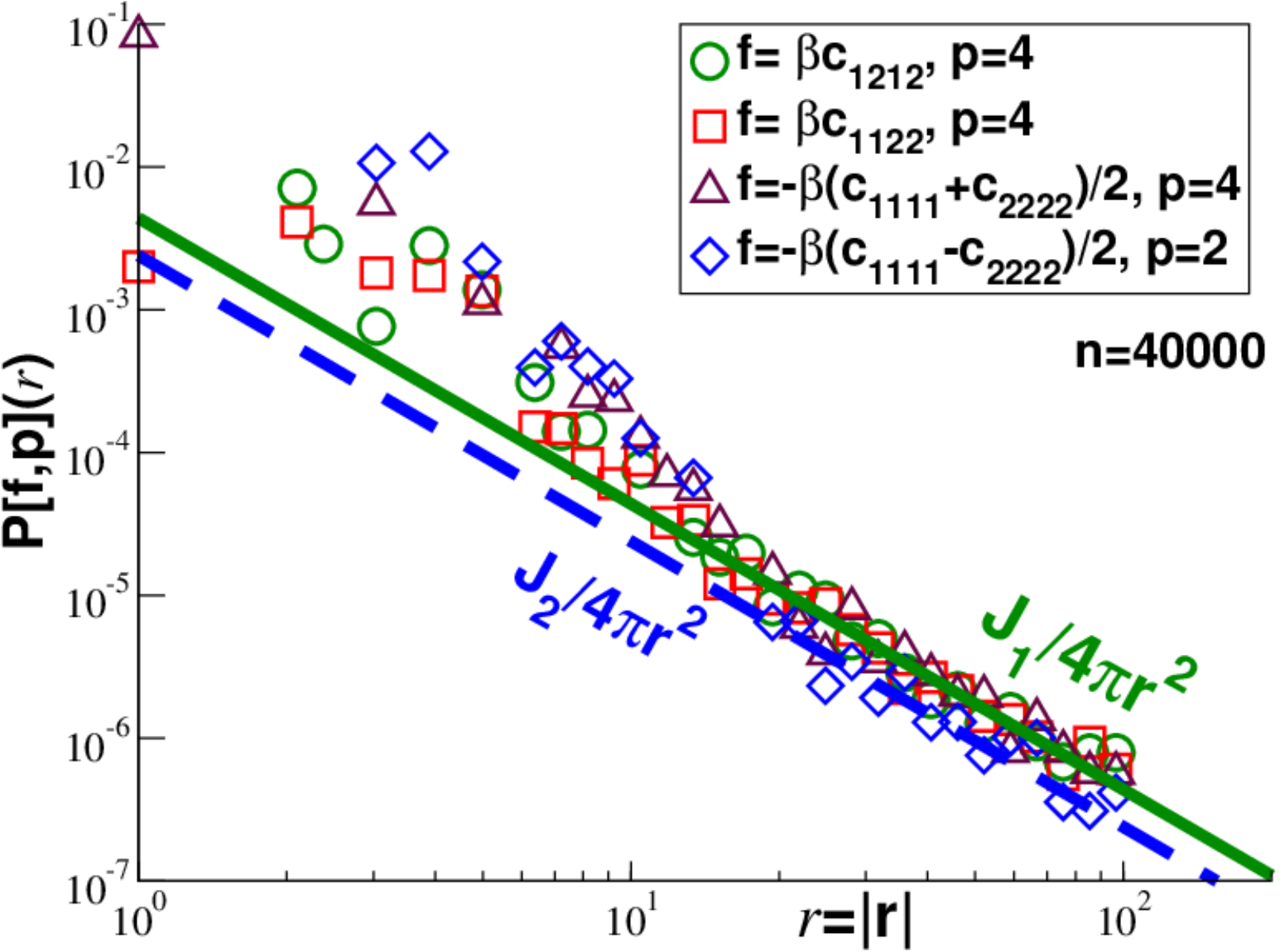}}}
\caption{$P[f,p](r)$ for various correlation functions and modes $p$ for $n=40000$.
The bold solid line marks the prediction $\Jone/4\pi r^2$ for the first three cases,
the dashed line the prediction $\Jtwo/4\pi r^2$ for the last one.
}
\label{fig_Cabcd_rproj}
\end{figure}

\subsection{Correlation functions in real space}
\label{res_Cabcd_r}

We turn finally to
the correlation functions $\cabcdprime(\rvec) = \Fcal^{-1}[\cabcdprime(\qvec)]$ in real space.
As shown in Appendix~\ref{app_Cabcd}, inverse FT implies 
\begin{equation}
\beta \cxyxyprime(\rvec) \simeq \frac{\Jone}{4\pi r^2} \cos(4x) \mbox{ for } r \gg 1
\label{eq_res_Cxyxy_rtheta}
\end{equation}
with $x=\theta-\alpha$ being the difference of the angles $\theta$ and $\alpha$ indicated in Fig.~\ref{fig_intro}.
The same large-$r$ limit holds also for  $\beta \cxxyyprime(\rvec)$ and for
$-\beta (\cxxxxprime(\rvec)+\cyyyyprime(\rvec))/2$.
Moreover,
\begin{equation}
\beta (\cxxxxprime(\rvec)-\cyyyyprime(\rvec))/2 \simeq - \frac{\Jtwo}{4\pi r^2} \cos(2x)
\label{eq_res_Cdiff_rtheta}
\end{equation}
for $r \gg 1$, i.e. a bi-polar symmetry is expected.
A verification of the $r$-dependence is obtained using again (now in real space) 
the projection $P[f,p](r)$, cf.~Eq.~(\ref{eq_res_Pfp_def}).
Focusing on $n=40000$ several rescaled correlation functions $f(\rvec)$ are presented in 
Fig.~\ref{fig_Cabcd_rproj}. In agreement with Eq.~(\ref{eq_res_Cxyxy_rtheta}) 
the indicated first three cases collapse for $p=4$ and $r \gtrsim 20$
on $\Jone/4\pi r^2$ (bold solid line). 
This confirms the octupolar symmetry 
of these correlation functions. Confirming Eq.~(\ref{eq_res_Cdiff_rtheta}) the last case with 
$f(\rvec)=-\beta (\cxxxx(\rvec)-\cyyyy(\rvec))/2$ collapses onto $\Jtwo/4\pi r^2$ (dashed line). 
$p=2$ is used here in agreement with the predicted quadrupolar
symmetry of this correlation function.
Similar results are obtained for other particle numbers $n$.

\section{Linear response to point stress}
\label{response}

\subsection{Time-dependent strain correlations}
\label{response_intro}

Correlation functions describe quite generally the linear response to a small 
imposed perturbation \cite{ChaikinBook,DoiEdwardsBook,ForsterBook,HansenBook}. 
Being tensorial fields, just like the correlation function fields, the response fields must in general 
depend on the direction of the field vector and on the orientation of the coordinate system.
As already emphasized in Sec.~\ref{theo_response} and Sec.~\ref{theo_source}, 
the response fields contains information of both the system and the imposed source 
and the source term may either be isotropic or anisotropic.
We elaborate here this general point focusing, naturally, on correlation functions of the instantaneous 
strain field $\epsabhat(\rvec)=\epsab(\rvec,t)$. 
Extending very briefly our discussion to the time domain,
let us introduce the time-dependent correlation functions
\begin{equation}
\cabcd(\qvec,t) = \la \epsab(\qvec,t) \epscd(-\qvec,t=0) \ra
\label{eq_cabcd_t_def}
\end{equation}
of the strain fields in reciprocal space with $t$ being the ``time lag" \cite{AllenTildesleyBook}. 
Naturally, this reduces to Eq.~(\ref{eq_intro_cabcd_def}) for $t\to 0$.
This definition allows us to take advantage of the general ``Fluctuation-dissipation theorem" (FDT) 
of statistical mechanics as stated, e.g., in Ref.~\cite{DoiEdwardsBook,ForsterBook,HansenBook}.
We thus anticipate immediate generalizations of the present study for time-dependent 
tensorial correlation and response fields which will be discussed elsewhere.

\subsection{Fluctuation-dissipation theorem}
\label{response_FDT}
Let us switch on at time $t=0$ a small perturbation 
\begin{equation}
\Delta \Hcal = -\int \ddiff \rvec \ \sigabimp(\rvec) \ \epsabhat(\rvec) \mbox{ for } t \ge 0
\label{eq_FDT_A}
\end{equation}
to the Hamiltonian $\Hcal=\Hcal_0+\Delta \Hcal$ of the system 
with $\sigabimp(\rvec)$ being an imposed external stress field.
This is equivalent to the application of an appropriate external perturbative force field to each particle.
For a general ``growth function" \cite{DoiEdwardsBook} in response to a sudden application of a step field, 
such as Eq.~(\ref{eq_FDT_A}), the relevant FDT relations are stated (for scalar fields) 
by Eq.~(3.65) and Eq.~(3.67) of Ref.~\cite{DoiEdwardsBook}.
The mean strain increment $\delta \epsab(\rvec,t)$ induced by this perturbation is then given in real space 
by a convolution integral for the time-dependent correlation functions $\cabcd(\rvec,t)$ and 
the stress perturbation $\sigabimp(\rvec)$.
Using Eq.~(\ref{eq_FT_convolution_q}) this relation may be written more compactly in reciprocal space as
\begin{eqnarray}
\delta \epsab(\qvec,t) & = & \beta V \label{eq_FDT_B} \\
&\times&[\cabcd(\qvec,t=0)-\cabcd(\qvec,t)] \sigcdimp(\qvec) \nonumber
\end{eqnarray}
where the summation over repeated indices is essential and cannot be omitted.
Note that $\delta \epsab(\qvec,t) = 0$ for $t \le 0$
and that, since all $\cabcd(\qvec,t)$ are continuous functions of time,
the creep response $\delta \epsab(\qvec,t)$ must also be continuous, especially at $t=0$ \cite{FerryBook}.
The time-dependent creep is thus determined by the time-dependent correlation functions
and the imposed stress perturbation.
We are interested here only in the static equilibrium response a long time 
after the perturbation is switched on. The time-dependent strain correlation functions 
(computed for the unperturbed Hamiltonian $\Hcal=\Hcal_0$ 
at switched off external perturbation $\Delta \Hcal$)
must, of course, vanish 
\begin{equation}
\cabcd(\qvec,t) \to 0 \mbox{ for } t \to \infty. 
\label{eq_FDT_C}
\end{equation}
Hence, Eq.~(\ref{eq_FDT_B}) reduces to 
\begin{equation}
\delta \epsab(\qvec) = \beta V \cabcd(\qvec) \sigcdimp(\qvec) \mbox{ for } t \to \infty
\label{eq_FDT_D}
\end{equation}
with $\delta \epsab(\qvec) \equiv \lim_{t\to\infty} \delta \epsab(\qvec,t)$ 
denoting the long-time creep 
and $\cabcd(\qvec) \equiv \lim_{t\to 0} \cabcd(\qvec,t)$ 
standing for the spatial correlation function without time lag, 
Eq.~(\ref{eq_intro_cabcd_def}), as everywhere else in this paper.

\subsection{Response to point source}
\label{response_source}

Following the discussion at the end of Sec.~\ref{theo},
we investigate now the long-time creep for a point source  
\begin{equation}
\sigabimp(\rvec) = s_{\alpha\beta} \delta(\rvec)
\label{eq_source_A}
\end{equation}
with $s_{\alpha\beta}$ being a symmetric $2 \times 2$ matrix (of dimension ``stress $\times$ volume = energy"). 
According to the FDT relation Eq.~(\ref{eq_FDT_D}) this implies
\begin{equation}
\delta \epsab(\qvec) = \beta \cabcd(\qvec) s_{\gamma\delta} \mbox{ for } t \to \infty
\label{eq_source_B}
\end{equation}
and an equivalent relation in real space.
As in Sec.~\ref{theo_response} it is convenient to diagonalize $s_{\alpha\beta}$ by an appropriate rotation 
of the coordinate system. The perturbation becomes therefore equivalent to that of two small force dipoles 
\cite{Bocquet04} oriented along the eigenvectors.
The real-space analogue of Eq.~(\ref{eq_theo_res_B}) is thus given by
\begin{equation}
\delta \epsab(\rvec) = 
\beta c_{\alpha\beta11}(\rvec) s_{11} + \beta c_{\alpha\beta22}(\rvec) s_{22}.
\label{eq_source_C}
\end{equation}
Using in addition Eq.~(\ref{eq_Cabcd_isomanifest}) we may replace for isotropic systems
the real space correlation functions by the corresponding invariants $\tilde{i}_n(r)$ in real space.
Introducing the scalars $s_1=s_{\gamma\gamma}/2$ and 
$s_2 = \rhat_{\alpha} s_{\alpha\beta} \rhat_{\beta}/2$ this may be written quite generally
\begin{eqnarray}
\delta \epsab(\rvec) & = & 
2 \left[\tilde{i}_1(r) s_1 + \tilde{i}_3(r) s_2\right] \delta_{\alpha\beta} \label{eq_source_D} \\
& + & 2 \left[\tilde{i}_3(r) s_1 + \tilde{i}_4(r) s_2\right] \rhat_{\alpha} \rhat_{\beta} \nonumber\\
&+ & 2 \tilde{i}_2(r) \ s_{\alpha\beta}. \nonumber
\end{eqnarray}
Taking now advantage of the specific results for strain correlations presented in Sec.~\ref{res_Cabcd_r} 
and in Appendix~\ref{app_Cabcd} the invariants $\tilde{i}_n(r)$ are given by Eq.~(\ref{eq_Cabcd_invariants}),
i.e. we may quite generally express $\delta \epsab(\rvec)$ in terms of 
the two creep compliances $\Jone$ and $\Jtwo$, cf.~Eq.~(\ref{eq_res_JoneJtwo}).

We also note that the term $s_{\alpha\beta}$ in the last line of Eq.~(\ref{eq_source_D}) 
must be isotropic, i.e. $s_1=s_{11}=s_{22}=2s_2$, to obtain an isotropic second-order tensor field
in agreement with Eq.~(\ref{eq_ten2_d2_r}). 
As expected from the more general argument given in Sec.~\ref{theo}, the shear strain increment 
\begin{equation}
\delta \varepsilon_{12}(\rvec)/s_{11} = - \frac{\Jtwo}{4\pi r^2} \sin(2\theta)
\mbox{ for } r > 0
\label{eq_source_epsxy_iso}
\end{equation}
becomes quadrupolar in this case.

As already emphasized in Sec.~\ref{theo_source}, the source tensor need not
necessarily be isotropic albeit the system is isotropic.
To be specific, let us consider the ``shear transformation zone" model
for localized plastic failure involving two orthogonal force dipoles of {\em opposite} signs 
\cite{Bocquet04}.
Hence, $s_{11}=-s_{22}$ and $s_1 =0$ and $s_2= s_{11} (\rhat_1^2-\rhat_2^2)/2$.
Using Eq.~(\ref{eq_theo_res_asosource}) or Eq.~(\ref{eq_source_D}) this yields 
\begin{equation}
\delta \varepsilon_{12}(\rvec)/s_{11}  
= - \frac{2 \Jone}{4\pi r^2} \sin(4\theta)
\mbox{ for } r > 0
\label{eq_source_epsxy_aniso}
\end{equation}
for the shear strain response.

As one expects on general grounds, all reference to statistical physics, 
i.e. the inverse temperature $\beta$, drops out in both cases.
Moreover, the shear strain response naturally strongly depends on the type of source term
applied at the origin: for force dipoles of same sign it is quadrupolar and proportional to $\Jtwo$ 
while for dipoles of opposite sign it gets octupolar and proportional to $\Jone$.

\section{Conclusion}
\label{conc}

\subsection{Summary}
\label{conc_summary}

\paragraph*{Strain correlation functions.}
The present work has focused on
correlation functions of components of strain tensor fields 
in two-dimensional, isotropic and achiral elastic bodies. 
This was done theoretically using 
\begin{itemize}
\item
the general mathematical structure of isotropic tensor fields 
as summarized in Sec.~\ref{theo_cabcd},
cf.~Eq.~(\ref{eq_theo_cabcd}), and in more detail in Appendix~\ref{ten} and 
\item
the well-known equipartition theorem of statistical physics for 
macroscopic strain fluctuations in reciprocal space, 
cf.~Eq.~(\ref{eq_intro_ICF_qlow}).
\end{itemize}
Numerically we have tested our predictions by means of glass-forming particles deep in the glass regime.
This shows that these correlation functions may depend on the coordinates of the field variable
($q_{\alpha}$ in reciprocal space or $r_{\alpha}$ in real space)
and implies in turn that they depend in general on the direction of the field vector and 
on the orientation of the coordinate system. 
Scaling with $x=\theta-\alpha$ these angular dependencies are distinct from those of ordinary anisotropic systems. 
Importantly, correlation functions of strain tensor fields are components 
of an isotropic forth-order tensor field, Eq.~(\ref{eq_theo_cabcd}), being characterized by the two ICFs
$\cone(q)$ and $\cfour(q)$.
With the asymptotic plateau values being given by two Lam\'e coefficients, 
Eq.~(\ref{eq_intro_ICF_qlow}), all strain correlation functions are determined
and all (finite) real-space strain correlations must be long-ranged decaying as $1/r^2$ 
(cf.~Fig.~\ref{fig_Cabcd_rproj}). 
We thus obtain similar results as in our recent study on correlation functions of stress tensor fields \cite{spmP5a}.
Note that {\em time-averaged} stress fields have been probed in the latter study while 
correlations of {\em instantaneous} strain fields have been considered here.
Our numerical findings do agree with other studies of strain
correlations \cite{Fuchs16,Fuchs18b,Reichman21b} being, however, now traced back to 
the isotropy of the system and the tensor field nature of the probed correlations.
Importantly, we have given here a complete and asymptotically exact description for the
correlation functions of strain tensor fields of isotropic elastic bodies. 
No additional physical assumption is thus needed (for sufficiently small wavevectors).
 
\paragraph*{Response to tensorial point sources.}
We also discussed the associated linear response fields
as defined in general mathematical terms by the tensorial contraction of the 
correlation function tensor by means of a source tensor 
and, more physically, by the FDT relation for the strain increment due to an imposed small
stress perturbation, cf.~Eq.~(\ref{eq_FDT_A}) and Eq.~(\ref{eq_FDT_B}). 
Naturally, the response must by definition contain information from both the correlation functions,
characterizing the system, and from the imposed source tensor which may not be isotropic.
We have emphasized that the summation over repeated indices must be properly performed,
i.e. the response field is {\em not} given by one correlation function times a scalar 
but by the sum over {\em all} eigenvalues of the source tensor. 
For this reason response and correlation fields, albeit closely related, have in general different
angular dependences, e.g., the shear strain correlation function $c_{1212}(\rvec)$ in an
isotropic system must be octupolar, cf.~Eq.~(\ref{eq_res_Cxyxy_rtheta}), while the
shear strain response $\delta \varepsilon_{12}(\rvec)$ may be either quadrupolar
for an isotropic source, cf.~Eq.~(\ref{eq_source_epsxy_iso}), or octupolar for
an anisotropic source corresponding to two force dipoles of opposite signs,
cf.~Eq.~(\ref{eq_source_epsxy_aniso}). 
Albeit all contributing correlation functions are isotropic
the response field is anisotropic in the latter case due to the source.
It is thus important to not lump together correlation functions and response fields.
Mesoscopic elasto-plastic models \cite{Tanguy11,Barrat18} thus must specify not only the correlation functions
but also the source tensors characterizing the local plastic events.

\subsection{Outlook}
\label{conc_outlook}

Our work suggests several natural extensions:
\begin{itemize}
\item
The general mathematical framework for isotropic tensor fields and the discussed relations and
numerical procedure for correlation and response fields naturally generalize to higher spatial dimensions, 
especially for the three-dimensional case.
\item
The present work has focused on Euclidean spaces and Cartesian coordinates.
A generalization for systems embedded in non-Euclidean spaces,
say for glasses on spheres \cite{Tarjus15,Tarjus17}, 
and more general curvilinear coordinate systems \cite{McConnell,Lambourne} may be worked out.
\item
The present work has focused on the large-wavelength limit ($q \to 0$).
More generally,
one may express the longitudinal and transverse ICFs $\cone(q)$ and $\cfour(q)$ for finite $q$ as
\begin{equation}
\beta V \cone(q)  = \frac{1}{L(q)} \mbox{ and } 
\beta V \cfour(q) = \frac{1}{4G(q)} \label{eq_outlook_equipartition}
\end{equation}
in terms of the generalized longitudinal and transverse elastic moduli $L(q)$ and $G(q)$ 
(with $L(q)\to \lambda+2\mu$ and $G(q) \to \mu$ for small $q$) 
\cite{lyuda18,lyuda21b,spmP5c}.\footnote{The elastic modulus tensor $\Eabcd(\qvec)$ for isotropic bodies
               at finite wavevector $\qvec$ is not only characterized by the
               longitudinal modulus $L(q)$ and the shear modulus $G(q)$
               but also by a third modulus $M(q)$ called the ``mixed modulus" \cite{lyuda18}.
               Note that $E_{1111}^{\circ}(q) = E_{2222}^{\circ}(q) = L(q)$,
               $E_{1212}^{\circ}(q) = G(q)$ and $E_{1122}^{\circ}(q) = M(q)$
               for isotropic bodies in NRC.
               It is not possible to determine $M(q)$ solely using strain fluctuations.
               This requires the additional measurement of stress fields in NRC.
               $M(q)$ may then be obtained using either the appropriate stress-strain or stress-stress
               correlation functions in reciprocal space \cite{lyuda18,spmP5c}.}
It can be shown \cite{spmP5c} that both the isotropicity and the harmonicity of the strain modes
assumed in the derivation of Eq.~(\ref{eq_outlook_equipartition}) are well justified for the present model 
up to $q \approx 1$ while deviations become relevant for larger wavevectors,
especially around the main peak of the static structure factor $S(q)$.
\item
A further generalization of the current work concerns time-dependent 
correlation functions $\cabcd(\qvec,t)$ as defined in Eq.~(\ref{eq_cabcd_t_def}).
These can be again expressed via Eq.~(\ref{eq_theo_cabcd}) in terms of 
(now time-dependent) longitudinal and transverse ICFs $\cone(q,t)$ and $\cfour(q,t)$.
These time-dependent ICFs are given in turn by time-dependent
creep compliance material functions which can be related to the two
time-dependent material functions $L(q,t)$ and $G(q,t)$ \cite{lyuda18}.
Strain correlation functions may thus reveal octupolar pattern whenever the invariant
\begin{equation}
|i_4(q,t)|=|\cone(q,t)-4 \cfour(q,t)|
\label{eq_outlook_invariant}
\end{equation} is sufficiently large.
Since $i_4(q,t)$ must become $q$-independent for small $q$, a long-range decay with
\begin{equation}
\cabcd(\rvec,t) \simeq 1/r^d
\label{eq_outllok_longrange}
\end{equation}
is generally expected for the time-dependent correlation functions in isotropic $d$-dimensional systems.
Using Eq.~(\ref{eq_FDT_B}) similar long-range relations are predicted for the associated dynamical response fields.
\item
It may be also of interest to characterize correlations of tensor fields of different order.
For instance, the forth-order elastic modulus field $E_{\alpha\beta\gamma\delta}(\rvec)$ 
\cite{Mossa19,Barrat13} may be characterized by a correlation function tensor of order eight \cite{Mossa19}.
Strong angular dependencies are expected based on our formalism.
For isotropic systems these correlation functions must again adopt a general mathematical structure
in terms of a small finite number of ICFs. 
Once these ICFs are characterized (theoretically or numerically using NRC)
in the low-$q$ limit all correlation functions are again determined.
\end{itemize}

\vspace*{0.2cm}
\begin{flushleft}
{\bf Author contributions:}
J.P.W. designed and wrote the project benefitting from contributions of all co-authors.

\vspace*{0.2cm}
{\bf Conflicts of interest:}
There are no conflicts to declare.

\vspace*{0.2cm}
{\bf Acknowledgments:}
We are indebted to the University of Strasbourg for computational re\-sources.
\end{flushleft}

\appendix
\section{Summary of isotropic tensor fields}
\label{ten}

\subsection{Background}
\label{ten_intro}

Isotropic systems are described by ``isotropic tensors" and ``isotropic tensor fields".
We give here a brief recap of various useful aspects already presented elsewhere \cite{Schultz_Piszachich,spmP5a}.
Quite generally, a tensor field assigns a tensor to each point of the mathematical space,
in our case a $d$-dimensional Euclidean vector space \cite{Schultz_Piszachich}.  
We denote an element of this vector space by the ``spatial position" $\rvec$ (real space) 
or by the ``wavevector" $\qvec$ for the corresponding reciprocal space.
The relations for tensor fields are formulated in reciprocal space since this is more 
convenient both on theoretical and numerical grounds due to the assumed spatial homogeneity
(``translational invariance"). 
The corresponding real space tensor field is finally obtained by inverse FT.
 
For simplicity we assume Cartesian coordinates with an orthonormal basis 
$\{\evec_1,\ldots,\evec_d\}$ \cite{McConnell,Schultz_Piszachich,TadmorCMTBook}.
Greek letters $\alpha, \beta,\ldots$ are used for the indices of the tensor (field) components. 
A twice repeated index $\alpha$ is summed over the values $1,\ldots,d$, e.g., 
$\qvec = q_{\alpha} \evec_{\alpha}$ with $q_{\alpha}$ standing for the vector coordinates. 
This work is chiefly concerned with tensors
$\Tvec^{(o)} = T_{\alpha_1\ldots\alpha_o} \evec_{\alpha_1} \ldots \evec_{\alpha_o}$
of ``order" $o=2$ and $o=4$ and their corresponding tensor fields 
with components depending either on $\rvec$ or $\qvec$. 
The order of a component is given by the number of indices.
Note that
\begin{equation}
T_{\alpha_1\ldots\alpha_o}(\qvec) = \Fcal[T_{\alpha_1\ldots\alpha_o}(\rvec)]
\label{eq_ten_intro_FT}
\end{equation}
for the $d^o$ coordinates in real and reciprocal space
(with $\Fcal[\ldots]$ denoting the FT as discussed in Appendix~\ref{app_FT}).
 
We consider linear orthogonal coordinate transformations (marked by ``$\Tgen$")
$\evec_{\alpha}^{\Tgen} = c_{\alpha\beta} \evec_{\beta}$ with matrix coefficients
$c_{\alpha\beta}$ given by the direction cosine 
$c_{\alpha\beta} \equiv \cos(\evec_{\alpha}^{\Tgen},\evec_{\beta})$ \cite{Schultz_Piszachich}.
We remind that \cite{Schultz_Piszachich}
\begin{equation}
T_{\alpha_1\ldots\alpha_o}^{\Tgen}(\qvec) =
c_{\alpha_1\nu_1} \ldots c_{\alpha_o\nu_o} T_{\nu_1\ldots\nu_o}(\qvec)
\label{eq_ten_intro_orthtrans}
\end{equation}
under a general orthogonal transform.
For a reflection of the $1$-axis we thus have, e.g.,
\begin{equation}
T_{1222}^{\Tgen}(\qvec) = -T_{1222}(\qvec),
T_{1221}^{\Tgen}(\qvec) = T_{1221}(\qvec),
\label{eq_ten_intro_refl1axis}
\end{equation}
i.e. we have sign inversion for an {\em odd number} of indices equal to the index of the inverted axis.
The field vector $\qvec=q_{\alpha}\evec_{\alpha}=q_{\alpha}^{\Tgen}\evec_{\alpha}^{\Tgen}$ 
remains unchanged by these ``passive" transforms albeit its coordinates change.

\subsection{Definitions, properties and construction of
general isotropic tensors and tensor fields}
\label{ten_iso}

\paragraph*{Isotropic tensors.}
Components of an isotropic tensor remain unchanged by {\em any} orthogonal coordinate transformation 
\cite{Schultz_Piszachich,TadmorCMTBook}, i.e.
\begin{equation}
T_{\alpha_1\ldots\alpha_o}^{\Tgen} =T_{\alpha_1\ldots\alpha_o}.
\label{eq_ten_iso}
\end{equation}
As noted at the end of Sec.~\ref{ten_intro} the sign of tensor components change 
for a reflection at one axis if the number of indices equal to the inverted axis is {\em odd}.
Consistency with Eq.~(\ref{eq_ten_iso}) implies that 
{\em all tensor components with an odd number of equal indices must vanish}, e.g., 
\begin{equation}
T_{12}=T_{1112} = T_{1222} = 0.
\label{eq_ten_iso_oddnumber_vanish}
\end{equation}

\paragraph*{Isotropic tensor fields.}
The corresponding isotropy condition for tensor fields is given by \cite{Schultz_Piszachich}
\begin{equation}
T_{\alpha_1\ldots\alpha_o}^{\Tgen}(q_1,\ldots,q_d) 
= T_{\alpha_1\ldots\alpha_o}(q_1^{\Tgen},\ldots,q_d^{\Tgen}) 
\label{eq_ten_iso_field}
\end{equation}
with $q_{\alpha}^{\Tgen} = c_{\alpha\beta}q_{\beta}$
which reduces to Eq.~(\ref{eq_ten_iso}) for $\qvec=\bfzero$.
Please note that the fields on the left handside of Eq.~(\ref{eq_ten_iso_field})
are evaluated with the original coordinates 
of the vector field variable $\qvec$ 
while the fields on the right handside are evaluated with the transformed coordinates.
Another way to state this is to say that the left hand fields are computed at
the original vector $\qvec=(q_1,\ldots,q_d)$ while the right hand fields are computed at
the ``actively transformed" vector $\qvec^{\Tgen} = (q_1^{\Tgen},\ldots,q_d^{\Tgen})$.
It is for this reason that finite components with an odd number of equal indices, e.g., 
$T_{1222}(\qvec) \ne 0$, are possible in principle for finite $\qvec$.

\paragraph*{Natural Rotated Coordinates.}
Fortunately, there is a convenient coordinate system where the nice symmetry Eq.~(\ref{eq_ten_iso_oddnumber_vanish}) 
for isotropic tensors can be also used for isotropic tensor fields. In these ``Natural Rotated Coordinates" (NRC)
the coordinate system for {\em each} wavevector $\qvec$ is rotated until the $1$-axis coincides with the 
$\qvec$-direction, i.e. $q_{\alpha}^{\circ} = q \delta_{1\alpha}$ with $q = |\qvec|$.
These tensor field components in NRC are marked by ``$\circ$" to distinguish them from standard rotated 
tensor fields (marked by primes ``$\prime$") where the {\em same} rotation is used for all $\qvec$.
If in addition $T_{\alpha_1\ldots\alpha_o}^{\circ}(\qvec)$ is an {\em even} function of its field variable $\qvec$
(as in the case of achiral systems for even order $o$) 
it can be shown \cite{spmP5a} that all tensor field components with an odd number of equal indices must vanish.

\paragraph*{Product theorem for isotropic tensor fields.}
Let us consider a tensor field 
$\Cvec(\qvec) = \Avec(\qvec) \otimes \Bvec(\qvec)$ 
with $\Avec(\qvec)$ and $\Bvec(\qvec)$ being two isotropic tensor fields
and $\otimes$ standing either for an outer product, 
e.g. $C_{\alpha\beta\gamma\delta}(\qvec)=A_{\alpha\beta}(\qvec)B_{\gamma\delta}(\qvec)$,
or an inner product, 
e.g. $C_{\alpha\beta\gamma\delta}(\qvec)=A_{\alpha\beta\gamma\nu}(\qvec)B_{\nu\delta}(\qvec)$.
Hence,
\begin{eqnarray}
\Cvec^{\Tgen}(\qvec) & = & \left(\Avec(\qvec) \otimes \Bvec(\qvec)\right)^{\Tgen} 
= \Avec^{\Tgen}(\qvec) \otimes \Bvec^{\Tgen}(\qvec) \nonumber \\
& = & \Avec(\qvec^{\Tgen}) \otimes \Bvec(\qvec^{\Tgen}) = \Cvec(\qvec^{\Tgen})
\label{eq_ten_iso_AB2C}
\end{eqnarray}
using in the second step a general property of tensor (field) products, due to 
Eq.~(\ref{eq_ten_intro_orthtrans}), and in the third step
Eq.~(\ref{eq_ten_iso_field}) for the fields $\Avec(\qvec)$ and $\Bvec(\qvec)$
where $\qvec^{\Tgen}$ stands for the ``actively" transformed field variable.
$\Cvec(\qvec)$ is thus also an isotropic tensor field.
One may use this theorem to construct
isotropic tensor fields from known isotropic tensor fields $\Avec(\qvec)$ and $\Bvec(\qvec)$. 

\subsection{Summary of assumed symmetries}
\label{ten_sym}

All second-order tensors in this work are symmetric, $T_{\alpha\beta}= T_{\beta\alpha}$,
and the same applies for the corresponding tensor fields in either $\rvec$- or $\qvec$-space.
This is, e.g., the case for the strain field $\epsab(\qvec) = \Fcal[\epsab(\rvec)]$, 
cf.~Eq.~(\ref{eq_algo_epsab}),
or the source tensor $s_{\alpha\beta}$ needed for a response field, cf.~Eq.~(\ref{eq_theo_res_A}).
We assume for all forth-order tensor fields that
\begin{eqnarray}
T_{\alpha\beta\gamma\delta}(\qvec) & = & T_{\beta\alpha\gamma\delta}(\qvec)=
T_{\alpha\beta\delta\gamma}(\qvec)
\label{eq_ten_sym_A}\\
T_{\alpha\beta\gamma\delta}(\qvec) & = & T_{\gamma\delta\alpha\beta}(\qvec) \mbox{ and } 
\label{eq_ten_sym_B} \\
T_{\alpha\beta\gamma\delta}(\qvec) & = & T_{\alpha\beta\gamma\delta}(-\qvec). \label{eq_ten_sym_C}
\end{eqnarray}
Note that Eq.~(\ref{eq_ten_sym_C}) is necessarily valid both for achiral and 
chiral two-dimensional isotropic systems.
Forth-order tensor fields are often constructed by taking
outer products \cite{TadmorCMTBook} of second-order tensor fields.
We consider, e.g., correlation functions
$\langle \hat{T}_{\alpha\beta}(\qvec) \hat{T}_{\gamma\delta}(-\qvec) \rangle$
with $\hat{T}_{\alpha\beta}(\qvec)$ being an instantaneous second-order tensor field. 
Eq.~(\ref{eq_ten_sym_A}) then follows from the symmetry of the second-order tensor fields.
The evenness of forth-order tensor fields,
Eq.~(\ref{eq_ten_sym_C}), is a necessary condition for {\em achiral} systems.
It implies that $T_{\alpha\beta\gamma\delta}(\qvec)$ is real if $T_{\alpha\beta\gamma\delta}(\rvec)$ 
is real and, moreover, Eq.~(\ref{eq_ten_sym_B}) for correlation functions since
\begin{equation}
\langle \hat{T}_{\alpha\beta}(\qvec) \hat{T}_{\gamma\delta}(-\qvec) \rangle
= \langle \hat{T}_{\gamma\delta}(\qvec) \hat{T}_{\alpha\beta}(-\qvec) \rangle.
\end{equation}
As already emphasized, all our systems are assumed to be {\em isotropic},
i.e., Eq.~(\ref{eq_ten_iso_field}) must hold for ensemble-averaged tensor fields.  

\subsection{General mathematical structure}
\label{ten_field}

\paragraph*{General structure of tensors.}
Isotropic tensors of different order are discussed, e.g., in Sec.~2.5.6 of Ref.~\cite{TadmorCMTBook}. 
Due to Eq.~(\ref{eq_ten_iso_oddnumber_vanish}) all such tensors of odd order must vanish. 
The finite isotropic tensors of lowest order are thus
\begin{eqnarray}
T_{\alpha\beta} & = & k_1 \delta_{\alpha\beta}, \label{eq_ten_ten_o2} \\
T_{\alpha\beta\gamma\delta} & = & i_1 \delta_{\alpha\beta} \delta_{\gamma\delta} 
+ i_2 \left(
\delta_{\alpha\gamma} \delta_{\beta\delta} + \delta_{\alpha\delta} \delta_{\beta\gamma}
\right) \label{eq_ten_ten_o4}
\end{eqnarray}
with $k_1$, $i_1$ and $i_2$ being invariant scalars.
Please note that all symmetries stated above hold,
especially also Eq.~(\ref{eq_ten_iso_oddnumber_vanish}).
Note that the symmetry Eq.~(\ref{eq_ten_sym_A}) was used for the second relation, Eq.~(\ref{eq_ten_ten_o4}).
Importantly, this implies that only {\em two} coefficients are needed for a forth-order isotropic tensor.
As a consequence, the elastic modulus tensor $\Eabcd$ is completely described by
{\em two} elastic moduli (cf. Sec.~\ref{app_comp_elastmacro}).

\paragraph*{General structure of tensor fields.}
We restate now the most general isotropic tensor fields for $1 \le o \le 4$
and focusing on two-dimensional systems ($d=2$) compatible with the symmetries stated in Sec.~\ref{ten_sym}.
With $l_n(q)$, $k_n(q)$, $j_n(q)$ and $i_n(q)$ being invariant scalar functions of $q=|\qvec|$
we have \cite{Schultz_Piszachich,spmP5a} 
\begin{eqnarray}
T_{\alpha}(\qvec) & = & l_1(q) \ \qhat_{\alpha} 
\label{eq_ten_field_o1} \\
T_{\alpha\beta}(\qvec) & = & 
k_1(q) \ \delta_{\alpha\beta} + k_2(q) \ \qhat_{\alpha}\qhat_{\beta} 
\label{eq_ten_field_o2} \\
T_{\alpha\beta\gamma}(\qvec) & = &
j_1(q) \ \qhat_{\alpha} \delta_{\beta\gamma} 
+ j_2(q) \ \qhat_{\beta}  \delta_{\alpha\gamma} \nonumber \\
& + & j_3(q) \ \qhat_{\gamma} \delta_{\alpha\beta} 
+ j_4(q) \ \qhat_{\alpha} \qhat_{\beta} \qhat_{\gamma}
\label{eq_ten_field_o3} \\
T_{\alpha\beta\gamma\delta}(\qvec) & = &
i_1(q) \ \delta_{\alpha\beta} \delta_{\gamma\delta} \label{eq_ten_field_o4} \\
& + & i_2(q) \ \left(
\delta_{\alpha\gamma} \delta_{\beta\delta} + \delta_{\alpha\delta} \delta_{\beta\gamma}
\right) \nonumber \\
& + & i_3(q) \ \left(
\qhat_{\alpha} \qhat_{\beta}\delta_{\gamma\delta} + \qhat_{\gamma}\qhat_{\delta}\delta_{\alpha\beta} 
\right) \nonumber \\
& + & i_4(q) \ \qhat_{\alpha} \qhat_{\beta} \qhat_{\gamma} \qhat_{\delta} \nonumber
\nonumber
\end{eqnarray}
for finite wavevectors $\qvec$.
See Ref.~\cite{spmP5a} for a derivation, generalizations for $d > 2$ and a discussion of the limit $q \to 0$.
Terms due to the invariants $k_1(q)$, $i_1(q)$ and $i_2(q)$ are independent of the coordinate system.
All other terms depend on the components $\qhat_{\alpha}$ of the normalized wavevector $\qhatvec$
and thus on the orientations of the field vector and of the coordinate system.

\paragraph*{Alternative representation for forth-order tensor fields.}
It is convenient to define the four functions
\begin{equation}
\left.
\begin{array}{ll}
\cone(q)   & \equiv T^{\circ}_{1111}(\qvec) \\
\cfour(q)  & \equiv T^{\circ}_{1212}(\qvec) \\
\ctwo(q)   & \equiv T^{\circ}_{2222}(\qvec) \\
\cthree(q) & \equiv T^{\circ}_{1122}(\qvec) 
\end{array}
\right\} \ \mbox{ for } \ q_{\alpha}^{\circ} = q \delta_{1\alpha} 
\label{eq_ten_Tabcd_d2_A} 
\end{equation}
using NRC.
For an isotropic system these four functions can only depend on the wavenumber $q$ 
but not on the direction $\hat{\qvec}$ of the wavevector $\qvec$. 
Importantly, all other components $T^{\circ}_{\alpha\beta\gamma\delta}(\qvec)$ 
are either by Eq.~(\ref{eq_ten_sym_A}) and Eq.~(\ref{eq_ten_sym_B}) identical to these invariants
or must vanish for an odd number of equal indices as reminded in Sec.~\ref{ten_iso}.
The $d^4=16$ components $T^{\circ}_{\alpha\beta\gamma\delta}(\qvec)$ are thus completely 
determined by the four invariants and this for any $\qvec$.
$T_{\alpha\beta\gamma\delta}(\qvec)$ is then obtained by the 
inverse rotation to the original unrotated frame using Eq.~(\ref{eq_ten_intro_orthtrans}). 
It is readily seen that
\begin{eqnarray}
\cone(q)   & = & i_1(q) + 2i_2(q) + 2i_3(q) + i_4(q) \nonumber \\
\ctwo(q)   & = & i_1(q) + 2i_2(q) \nonumber \\
\cthree(q) & = & i_1(q) + i_3(q) \nonumber \\
\cfour(q)  & = & i_2(q) 
\label{eq_ten_Tabcd_d2_B} 
\end{eqnarray}
being consistent with Eq.~(\ref{eq_theo_in2cn}).

\paragraph*{Isotropic tensor fields in real space.}
We have formulated above all tensor fields in terms of the wavevector $\qvec$ and its components 
since it is most convenient to start the analysis in reciprocal space.
The above results also hold, however, in real space.
This implies, e.g., for isotropic (and achiral) fields in two dimensions that 
\begin{eqnarray}
T_{\alpha\beta}(\rvec) & = &
\tilde{k}_1(r) \ \delta_{\alpha\beta} + \tilde{k}_2(r) \ \rhat_{\alpha} \rhat_{\beta}
\label{eq_ten2_d2_r} \\
T_{\alpha\beta\gamma\delta}(\rvec) & = &
\tilde{i}_1(r) \ \delta_{\alpha\beta} \delta_{\gamma\delta} \label{eq_ten4_d2_r} \\
& + & \tilde{i}_2(r) \left[
\delta_{\alpha\gamma} \delta_{\beta\delta} + \delta_{\alpha\delta} \delta_{\beta\gamma}
\right] \nonumber \\
& + & \tilde{i}_3(r) \left[
\rhat_{\alpha} \rhat_{\beta}\delta_{\gamma\delta} + \rhat_{\gamma} \rhat_{\delta}\delta_{\alpha\beta} 
\right] \nonumber \\
& + & \tilde{i}_4(r) \ \rhat_{\alpha} \rhat_{\beta} \rhat_{\gamma} \rhat_{\delta} \nonumber
\end{eqnarray}
for $r>0$ with $\tilde{k}_n(r)$ and  $\tilde{i}_n(r)$ denoting the invariants in real space
and $\rhat_{\alpha}=r_{\alpha}/r$ components of the normalized vector $\rhatvec=\rvec/r$.
As already stated, Eq.~(\ref{eq_ten_intro_FT}), the tensor field components
in real and reciprocal space are related by FT.
Note that $\tilde{k}_n(r)$ and $\tilde{i}_n(r)$ are in general {\em not} the FTs 
of, respectively, $k_n(q)$ and  $i_n(q)$.
For the important case that the invariants in reciprocal space are $q$-independent constants
it follows quite generally that
\begin{eqnarray}
4\pi r^2 \ \tilde{k}_1(r) & = & 2 k_2          \label{eq_ten_d2_intilde} \\
4\pi r^2 \ \tilde{k}_2(r) & = & -4 k_2         \nonumber \\
4\pi r^2 \ \tilde{i}_1(r) & = & 4 i_3 + 5 i_4  \nonumber \\
4\pi r^2 \ \tilde{i}_2(r) & = & - i_4          \nonumber \\ 
4\pi r^2 \ \tilde{i}_3(r) & = & -4 i_3 - 6 i_4 \nonumber \\ 
4\pi r^2 \ \tilde{i}_4(r) & = & 8 i_4          \nonumber  
\end{eqnarray}
for $r > 0$.
(Additional $\delta(\rvec)$-terms arise at the origin. 
The constant invariants $k_1$, $i_1$ and $i_2$ only contribute to these terms.)
That this holds can be readily shown using relations put forward
in Appendix~\ref{app_FT} and Appendix~\ref{app_Cabcd}. 

\section{Useful Fourier transformations}
\label{app_FT}

\subsection{Continuous Fourier transform}
\label{app_FT_continuous}

We consider real-valued functions $f(\rvec)$ in $d$ dimensions. 
As in Refs.~\cite{spmP4,lyuda22a,spmP5a} the Fourier transform (FT) $f(\qvec) = \Fcal[f(\rvec)]$ 
from ``real space" (variable $\rvec$) to ``reciprocal space" (variable $\qvec)$ is defined by
\begin{equation}
f(\qvec) = \frac{1}{V} \int \ddiff \rvec \ f(\rvec) \exp(-i \qvec \cdot \rvec)
\label{eq_FT_def}
\end{equation}
with $V$ being the volume of the system. The inverse FT is then given by
\begin{equation}
f(\rvec) = \Fcal^{-1}[f(\qvec)] 
= \frac{V}{(2\pi)^d} \int \ddiff \qvec \ f(\qvec) \exp(i \qvec \cdot \rvec).
\label{eq_FT_def_inv}
\end{equation}
Note that $f(\rvec)$ and $f(\qvec)$ have the same dimension.
For notational simplicity the function names remain unchanged.
We remind the FTs
\begin{eqnarray}
\Fcal\left[\frac{\partial}{\partial  r_{\alpha}} f(\rvec)\right] & = & i q_{\alpha} f(\qvec) \label{eq_FT_A}\\
\Fcal\left[\delta(\rvec-\vvec)\right] & = & \frac{1}{V} \exp(-i \qvec \cdot \vvec) \label{eq_FT_B} 
\end{eqnarray}
with $\delta(\rvec)$ being Dirac's delta function.
Let us consider the spatial convolution function
\begin{equation}
f(\rvec) = \frac{1}{V} \int \ddiff \rvec' g(\rvec-\rvec') h(\rvec')
\label{eq_FT_convolution_r} 
\end{equation}
in real space. With $g(\qvec) = \Fcal[g(\rvec)]$ and $h(\qvec) = \Fcal[h(\rvec)]$
this implies according to the ``convolution theorem" \cite{numrec}
\begin{equation}
f(\qvec)  = \Fcal[f(\rvec)] = g(\qvec) h(\qvec).
\label{eq_FT_convolution_q} 
\end{equation}
We also remind for completeness that the spatial correlation function
\begin{equation}
c(\rvec) = \frac{1}{V} \int \ddiff \rvec' g(\rvec+\rvec') h(\rvec')
\label{eq_FT_correlation_r} 
\end{equation}
of real-valued fields $g(\rvec)$ and $h(\rvec)$ becomes according to the 
``correlation theorem" \cite{numrec}
\begin{equation}
c(\qvec)  = g(\qvec) h^{\star}(\qvec) =  g(\qvec) h(-\qvec)
\label{eq_FT_correlation_q} 
\end{equation}
with $\star$ marking the conjugate complex.
For auto-correlation functions, i.e. for $g(\rvec)=h(\rvec)$,
this simplifies to (``Wiener-Khinchin theorem") 
\begin{equation}
c(\qvec)  = g(\qvec) g^{\star}(\qvec) = |g(\qvec)|^2,
\label{eq_FT_WKT} 
\end{equation}
i.e. the Fourier transformed auto-correlation functions are real and $\ge 0$ for all $\qvec$.
Moreover, we shall consider correlation functions $c(\rvec)$, 
Eq.~(\ref{eq_FT_correlation_r}), being {\em even} in real space, $c(\rvec)=c(-\rvec)$, 
and thus also in reciprocal space, $c(\qvec)=c(-\qvec)=c^{\star}(\qvec)$,
i.e. $c(\qvec)$ is real. 

\subsection{Discrete Fourier transform on microcell grid}
\label{app_FT_grid}

All fields $f(\rvec)$ are stored on a regular equidistant $d$-dimensional 
grid as shown in Fig.~\ref{fig_grid} for $d=2$. Periodic boundary conditions 
are assumed \cite{AllenTildesleyBook}. 
The discrete FT and its inverse become
\begin{eqnarray}
f(\qvec) & = & \frac{1}{\nVgrid} \sum_{\rvec} f(\rvec) \exp(-i \qvec\cdot \rvec)
\label{eq_FT_discrete} \\
f(\rvec) & = & \sum_{\qvec} f(\qvec) \exp(i \qvec\cdot \rvec)
\label{eq_FT_discrete_inv} 
\end{eqnarray}
with $\sum_{\rvec}$ and $\sum_{\qvec}$ being discrete sums over $\nVgrid=\nLgrid^d=V/\agrid^d$ grid points in,
respectively, real or reciprocal space. 
As shown in Fig.~\ref{fig_grid} we label the grid points in real and reciprocal space using
\begin{eqnarray}
\frac{r_{\alpha}}{\agrid} & = & n_{\alpha} \mbox{ and } q_{\alpha} \agrid = \frac{2\pi}{\nLgrid} n_{\alpha}
\label{eq_FT_gridpoints} \\
\mbox{with }
n_{\alpha} & = &-\frac{\nLgrid}{2}+1,\ldots,0,1,\ldots, \frac{\nLgrid}{2}. \nonumber
\end{eqnarray}
To take advantage of the Fast-Fourier transform (FFT) routines \cite{numrec} 
the number of grid points in each spatial direction $\nLgrid=L/\agrid$ is an integer-power of $2$.

\subsection{Fourier transform of planar harmonic functions}
\label{app_FT_inverse}

As discussed in the main part, all correlation functions in reciprocal space become in the 
large-wavelength limit independent of
the magnitude $q$ of the wavevector $\qvec$ but depend on its angle $\thetaq$ 
(and, more generally, on the angle difference $\thetaq-\alpha$ for rotated coordinate frames).
As noted in Sec.~\ref{theo_harmonics}, these angular dependencies can be uniquely 
expressed in terms of the planar harmonic basis functions $\cos(p \thetaq)$ and $\sin(p \thetaq)$ 
with $p$ being an integer. We denote these orthogonal basis functions by $b_p(\thetaq)$.
We thus need to compute the inverse FTs of $f(\qvec) = b_{p}(\thetaq)/V$.
More specifically, we are interested in modes with $p=2$ and $p=4$.
Additional constant terms ($p=0$), such as the ones indicated 
by dots in Eqs.~(\ref{eq_res_cxyxy_q}-\ref{eq_res_cmean_q}),
are irrelevant leading merely to $\delta(\rvec)$-contributions at the origin.
For $d=2$ Eq.~(\ref{eq_FT_def_inv}) becomes 
\begin{eqnarray}
f(\rvec) 
 & = & \frac{1}{4\pi^2} \int_0^{\infty} \ddiff q \ q \times \nonumber \\
 & & \hspace*{-.5cm} \int_0^{2\pi} \ddiff \thetaq \ b_p(\thetaq) \exp[i q r \cos(\thetaq-\thetar)]
\label{eq_q2r_invFT}
\end{eqnarray}
with $\thetar$ being the angle of $\rhatvec = (\cos(\thetar),\sin(\thetar))$.
We make now the substitution $\theta = \thetaq-\thetar$ and use that \cite{abramowitz}
\begin{eqnarray}
\cos(p\theta+p\thetar)+\cos(-p\theta+p\thetar) & = & 2 \cos(p\theta) \cos(p\thetar) 
\nonumber \\ 
\sin(p\theta+p\thetar)+\sin(-p\theta+p\thetar) & = & 2 \cos(p\theta) \sin(p\thetar).
\nonumber
\end{eqnarray}
We remind that following Eq.~(9.1.21) of Ref.~\cite{abramowitz} 
the integer Bessel function $J_p(z)$ may be written
\begin{equation}
J_p(z) = \frac{i^{-p}}{\pi} \int_0^{\pi} \ddiff \theta \cos(p\theta) \exp[i z \cos(\theta)]
\label{eq_q2r_Jp}
\end{equation}
which leads to
\begin{equation}
f(\rvec) = \frac{i^p}{2\pi} b_p(\thetar) \ \int_0^{\infty} \ddiff q \ q \ J_p(rq). 
\label{eq_q2r_JpB}
\end{equation}
For finite $r$ we may rewrite this as
\begin{equation}
f(\rvec) = \frac{i^p p}{2\pi r^2} b_p(\thetar) \times \lim_{t\to \infty} I_p(t)
\mbox{ for } r > 0
\label{eq_q2r_JpBB}
\end{equation}
where we have set $I_p(t) \equiv \int_0^t \ddiff t' t' J_p(t')/p$.
As may be seen from Eq.~(11.4.16) of Ref.~\cite{abramowitz} the latter integral 
becomes\footnote{The infinite integral Eq.~(11.4.16) of Ref.~\cite{abramowitz}
            is expressed in terms of two coefficients $\mu$ and $\nu$
            which in our case take the (real) values $\mu=1$ and $\nu=p$.
            It is stated that Eq.~(11.4.16) holds for $\Re(\mu+\nu) > -1$,
            being consistent with the condition $p > -2$ noted in Eq.~(\ref{eq_q2r_JpC}),
            but also that $\Re \mu < 1/2$, being at first sight in conflict with $\mu=1$.
            However, the integral divergence for $t \to \infty$ for $\mu>1/2$ is fictitious as,
            e.g., discussed in the Wikipedia entry on ``Oscillatory integrals"
            where it is noted that {\em "Oscillatory integrals make rigorous many arguments that,
            on a naive level, appear to use divergent integrals."}
            Note that Eq.~(\ref{eq_q2r_fr_p}) also holds for $p=0$ and
            finite $r>0$ which follows from the fact that the FT of
            a constant is zero everywhere except at the origin.
            See Ref.~\cite{spmP5a} for an alternative more straight-forward but also
            more lengthy derivation of Eq.~(\ref{eq_q2r_fr_p}) using the asymptotic behavior of
            the confluent hypergeometric Kummer function $M(a,b,z)$.}
\begin{equation}
I_p(t) \to 1 \mbox{ for } t \to \infty \mbox{ and } p > -2
\label{eq_q2r_JpC}
\end{equation}
from which we obtain the final result
\begin{equation}
f(\rvec) =  \frac{i^p p}{2\pi r^2} \ b_{p}(\thetar) 
\mbox{ for } r > 0.
\label{eq_q2r_fr_p}
\end{equation}
Note that $f(\rvec)=f(-\rvec)$ and that $f(\rvec)$ is real for even $p$.
Generalizing the above argument it is seen that $f(\rvec) \propto 1/r^d$ for higher dimensions $d$.

\section{Computational details}
\label{app_comp}

\subsection{Simulation model}
\label{app_comp_model}

We consider systems of polydisperse Lennard-Jones (pLJ) particles in $d=2$ dimensions
where two particles $i$ and $j$ of diameter $D_i$ and $D_j$ interact by means of a
central pair potential \cite{WXP13,lyuda19a,spmP1,spmP2,spmP5a,lyuda21b,lyuda22a} 
\begin{equation}
u(s) = 4 \epsilon \left(\frac{1}{s^{12}}-\frac{1}{s^{6}}\right)
\mbox{ with } s=\frac{r}{(D_i+D_j)/2}
\label{eq_comp_Us}
\end{equation}
being the reduced distance according to the Lorentz rule \cite{HansenBook}.
This potential is truncated and shifted \cite{spmP5a,AllenTildesleyBook}
with a cutoff $\scut=2\smin$ given by the minimum $\smin$ of $u(s)$.
Lennard-Jones units \cite{AllenTildesleyBook} are used throughout this study,
i.e. $\epsilon=1$ and the average particle diameter is set to unity.
The diameters are uniformly distributed between $0.8$ and $1.2$.
We also set Boltzmann's constant $\kB=1$ and assume that all particles have the same mass $m=1$. 
The last point is irrelevant for the presented Monte Carlo (MC) simulations
\cite{AllenTildesleyBook}.
Time is measured in units of MC steps (MCS) throughout this work.

\subsection{Parameters and configuration ensembles}
\label{app_comp_config}

We focus on systems with $n=10000$ and $n=40000$ particles.
We first equilibrate $\Nc=200$ independent configurations $c$ at a high temperature $T=0.55$ in the liquid limit.
These configurations are adiabatically cooled down using a combination of local MC moves
\cite{AllenTildesleyBook} and swap MC moves exchanging the sizes of pairs of particles \cite{Berthier17,spmP1}.
In addition, an MC barostat \cite{AllenTildesleyBook} imposes an average normal stress $P=2$ \cite{WXP13,spmP1}.
At the working temperature $T=0.2$ we first thoroughly temper over $\tsamp=10^7$ 
all configurations with switched-on local, swap and barostat MC moves
and then again over $\tsamp=10^7$ with switched-on local and swap moves and switched-off barostat moves.
The final production runs are carried out at constant volume $V$ only keeping local MC moves.
Under these conditions, $T=0.2$ is well below the glass transition temperature $\Tglass \approx 0.26$ 
determined in previous work \cite{WXP13,spmP1}.
Due to the barostat used for the cooling
the box volume $V=L^d$ differs slightly between different configurations $c$
while $V$ is identical for all frames $t$ of the time-series
of the same configuration $c$.
In all cases the number density is of order unity. 
For each particle number $n$ and each of the $\Nc$ independent configurations $c$
we store ensembles of time series containing $\Nt=10000$ instantaneous ``frames" $t$.
These are obtained using the equidistant time intervals $\tincr=1000$ for $n=10000$
and $\tincr=100$ for the other system sizes.

\subsection{Macroscopic linear elastic properties}
\label{app_comp_elastmacro}

The amorphous glasses formed by pLJ particle systems at a pressure $P=2$ and a temperature
$T=0.2 \ll \Tglass \approx 0.26$ are for sampling (production) times $\tsamp \le 10^{7}$ MCS
reversible {\em linear elastic bodies} whose plastic rearrangements can be neglected 
for all practical purposes \cite{WXP13,WXP13c,WXB15,WXBB15,spmP1,lyuda22a}.
Moreover, these systems can be shown to be {\em isotropic} above distances corresponding to a couple
of particle diameters \cite{spmP1,lyuda22a}.
Following Eq.~(\ref{eq_ten_ten_o4}) the forth-order elastic modulus tensor $\Eabcd$
for isotropic systems may be written \cite{LandauElasticity,TadmorCMTBook} 
\begin{equation}
\Eabcd = \lambda \delta_{\alpha\beta} \delta_{\gamma\delta}
+ \mu \left(\delta_{\alpha\gamma}\delta_{\beta\delta}+
\delta_{\alpha\delta}\delta_{\beta\gamma}\right)
\label{eq_elastmacro_Eabcd_iso}
\end{equation}
in terms of the two isothermic Lam\'e moduli $\lambda$ and $\mu$.
As described in detail elsewhere \cite{Lutsko88,Lutsko89,WXP13,WXP13c,WXB15,WXBB15}
we have determined $\lambda$ and $\mu$ either by means of strain fluctuations,
e.g., by letting the box volume $V$ fluctuate at imposed pressure $P$ \cite{WXP13c},
or using the stress-fluctuation formalism at fixed volume and shape of the simulation box
\cite{Lutsko88,Lutsko89}. 
This shows that $\lambda \approx 38$ and $\mu \approx 14$.
We have verified especially that similar values are obtained for $n \ge 5000$
and using different components, say $E_{1111}$ and $E_{2222}$ for $\lambda + 2\mu$,
and that the fluctuations of $\Eabcd|_c$ for independent configurations $c$ become
negligible for $n \ge 5000$.

\subsection{Discrete fields on square grid}
\label{app_comp_grid}

\begin{figure}[t]
\centerline{\resizebox{0.9\columnwidth}{!}{\includegraphics*{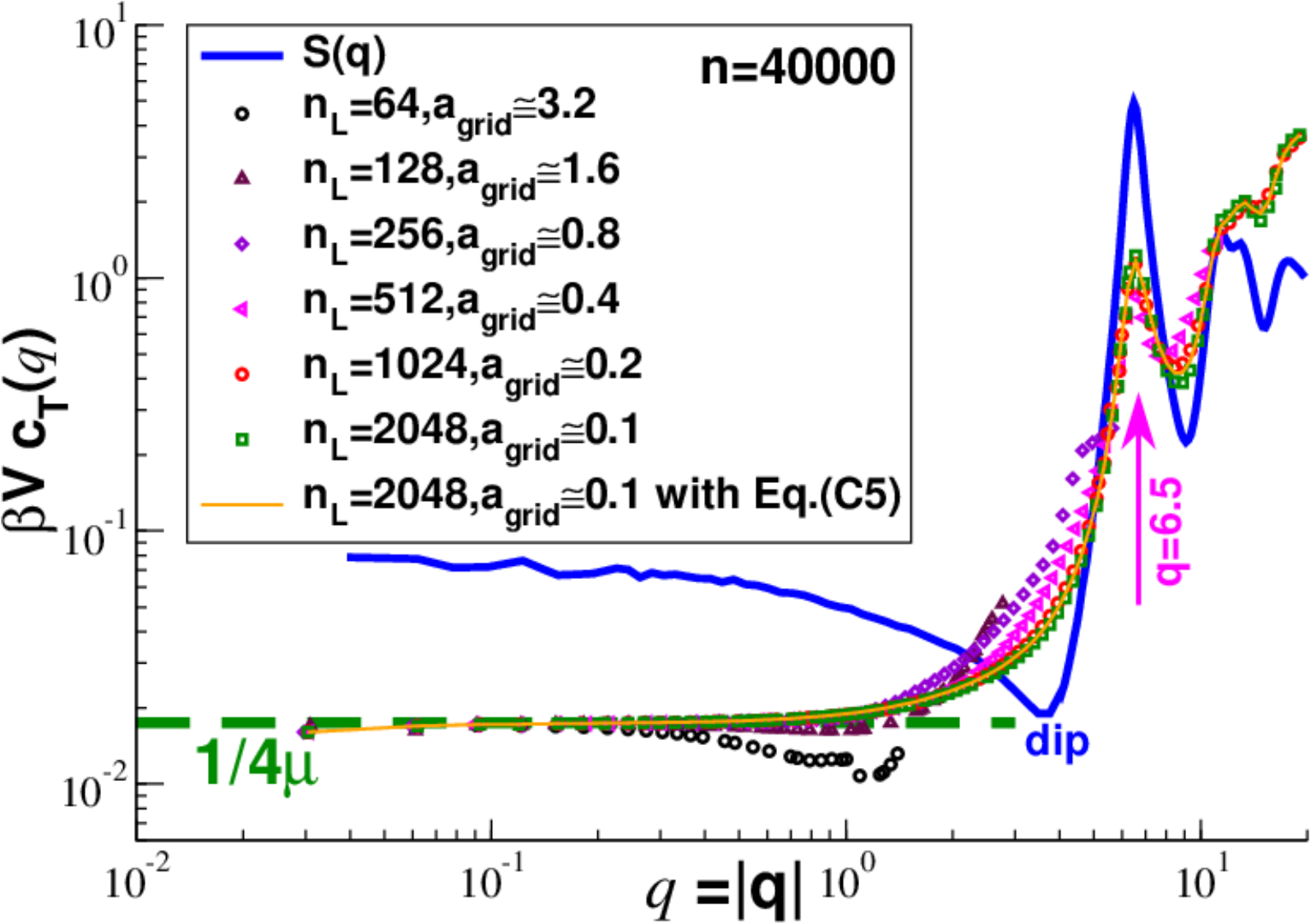}}}
\caption{Rescaled transverse ICF $\beta V \cfour(q)$ 
for different grid constants $\agrid$ as indicated.
The open symbols have been obtained using Eq.~(\ref{eq_uq_ur_def_A}),
the thin solid line using Eq.~(\ref{eq_uq_ur_def_B}) and $\nLgrid=2048$.
Importantly, we obtain the same results in all cases 
where $q \urmsq \ll 1$ and $q \agrid \ll 1$.
Even a rather coarse grid, say for $\nLgrid=64$, is sufficient to confirm the
expected large-wavelength limit (horizontal dashed line).
The total static structure factor $S(q)$ is shown for comparison (solid line).
The ``dip" of $S(q)$ around $q \approx 4$ is caused by the polydispersity of the particles
as emphasized elsewhere \cite{lyuda21b}. 
$S(q)$ and $\beta V \cfour(q)$, at least for sufficiently small $\agrid$, have both a strong peak
located similarly at $q \approx 6.5$ (arrow).
}
\label{fig_comp_uq}
\end{figure}

We turn now to the relevant microscopic tensor fields as functions of either 
the spatial position $\rvec$ (real space) or the wavevector $\qvec$ (reciprocal space).
The different fields are stored on equidistant discrete grids as sketched in Fig.~\ref{fig_grid}.
The same $\nLgrid$ is used for both spatial directions and for all configurations and frames 
of a given particle number $n$. 
As already mentioned, the box volume $V=L^2$ fluctuates slightly between different configurations $c$ 
(at same $n$) due to the barostat used for the cooling, tempering and equilibration of the systems. 
Accordingly, $\agrid$ also differs between different configurations $c$.
These fluctuations become small, however, with increasing system size.
If nothing else is mentioned we report data obtained 
using a lattice constant $\agrid \approx 0.2$.
As shown in Fig.~\ref{fig_comp_uq} for the rescaled transverse ICF $\beta V \cfour(q)$
plotted using a double-logarithmic representation, there is no need to further decrease $\agrid$.
Even the very large grid constant $\agrid \approx 3.2$ gives, apparently, the correct large-wavelength 
asymptote $\beta \cfour(q) \approx 1/4\mu$ indicated by the dashed horizontal line. 

\subsection{Displacement fields}
\label{app_comp_uq}

As in previous experimental and numerical studies \cite{Klix12,Klix15,WXBB15} the displacement field
$\uvec(\rvec)$ is constructed from the instantaneous spatial positions $\rvec_a$ of the 
particles $a$ with respect to their reference positions $\rref_a$. 
As reference position $\rref_a$ we have used either the average particle position determined using a long trajectory
or the particle position after a rapid quench to $T=0$. Having not observed any significant  
quantitative difference between both methods we only report here data computed using the first one.
We thus get first the displacement vector $\uvec_a = \rvec_a - \rref_a$ for each $a$.
By construction the average displacement vector $\la \uvec_a \ra$ must vanish. We find 
\begin{equation}
\urmsq \equiv \la \uvec_a^2 \ra^{1/2} \approx 0.13
\label{eq_uq_urmsq}
\end{equation}
for the root-mean-squared average $\urmsq$
(sampled over all particles, frames and configurations).
The instantaneous displacement field may then be defined by \cite{Klix12,Klix15,WXBB15}
\begin{equation}
\uvec(\rvec) = \frac{1}{n/V} \sum_a \uvec_a \delta(\rvec-\rref_a)
\label{eq_uq_ur_def_A}
\end{equation}
assuming for the moment an infinitessimal fine grid, i.e. $\agrid \to 0$.
(Different prefactors have been used in Refs.~\cite{Klix12,Klix15,WXBB15}.)
We remind that the $\delta(\rvec)$-function has the dimension ``$1/$volume".
By definition $\uvec(\rvec)$ has thus the same dimension ``length" as the displacement vector $\uvec_a$.
Following the common definition of the particle flux density \cite{HansenBook},
the reference position $\rref_a$ in the $\delta$-function may be replaced by the
time-dependent position $\rvec_a$, i.e. the displacement field may alternatively be defined by
\begin{equation}
\uvec(\rvec) = \frac{1}{n/V} \sum_a \uvec_a \delta(\rvec-\rvec_a).
\label{eq_uq_ur_def_B}
\end{equation}
Both operational definitions are compared for the transverse ICF $\beta V \cfour(q)$ 
in Fig.~\ref{fig_comp_uq} where data obtained using Eq.~(\ref{eq_uq_ur_def_A}) are 
indicated by open symbols. In reciprocal space we obtain 
\begin{equation}
\uvec(\qvec) = \frac{1}{n} \sum_a \uvec_a \exp(-i \qvec \cdot \rref_a).
\label{eq_uq_uq_def}
\end{equation}
for Eq.~(\ref{eq_uq_ur_def_A}) and similarly for Eq.~(\ref{eq_uq_ur_def_B}) with $\rvec_a$ replacing $\rref_a$.
Since $\rvec_a = \rref_a + \uvec_a$ we have to leading order
\begin{equation}
\exp(-i \qvec \cdot \rvec_a) \approx \exp(-i \qvec \cdot \rref_a) \ (1-i \qvec \cdot \uvec_a \ldots)
\label{eq_uq_expansion}
\end{equation}
for $q |\uvec(\qvec)| \ll 1$. Both operational definitions Eq.~(\ref{eq_uq_ur_def_A}) and 
Eq.~(\ref{eq_uq_ur_def_B}) thus become equivalent for $q \urmsq \ll 1$.
Due to the small typical (root-mean-squared) displacement $\urmsq$, Eq.~(\ref{eq_uq_urmsq}),
this holds for all sampled $q$ as may be seen from the data presented in Fig.~\ref{fig_comp_uq}.
Due to the center-of-mass convention for all particle displacements
the volume integral over $\uvec(\rvec)$ must vanish
and, equivalently, we have $\uvec(\qvec=0)=0$ in reciprocal space for each instantaneous field.
We also remind that the two coordinates of the displacement field in NRC are the
longitudinal component $\ulongi(\qvec) \equiv u_1^{\circ}(\qvec)$ and
the transverse component $\utrans(\qvec) \equiv u_2^{\circ}(\qvec)$.
 
In numerical practice, the continuous field vector $\rvec$ of 
Eq.~(\ref{eq_uq_ur_def_A}) and Eq.~(\ref{eq_uq_ur_def_B}) 
corresponds to the discrete point on the grid, Eq.~(\ref{eq_FT_gridpoints}),
closest (using the minimal image convention) to, respectively,
the reference position $\rref_a$ or the particle position $\rvec_a$.
Strictly speaking, we thus obtain by means of Eq.~(\ref{eq_FT_discrete}) 
the FT with respect to their respective closest grid points.
(In principle one could directly without approximation compute the displacement field $\uvec(\qvec)$ 
using Eq.~(\ref{eq_uq_uq_def}) in reciprocal space.
Unfortunately, this leads to an additional loop over all $n$ particles for each wavevector $\qvec$.)
The differences between these definitions become neglible for $q \agrid \ll 1$.
That this holds can be clearly seen from Fig.~\ref{fig_comp_uq} where
we have varied $\agrid$ over more than one order of magnitude.

\subsection{Linear strain fields}
\label{app_comp_strain}

Using the displacement field $\uvec(\rvec)$ the linear (``small") strain tensor field is defined by 
\cite{LandauElasticity,TadmorCMTBook}
\begin{equation}
\epsab(\rvec) \equiv \frac{1}{2}
\left(
\frac{\partial u_{\alpha}(\rvec)}{\partial r_{\beta}} + 
\frac{\partial u_{\beta}(\rvec)}{\partial r_{\alpha}}
\right).
\label{eq_comp_strain_r}
\end{equation}
Due to Eq.~(\ref{eq_FT_A}) this becomes 
\begin{equation}
\epsab(\qvec) =
\frac{i}{2}
\left(q_{\beta} u_{\alpha}(\qvec) + q_{\alpha} u_{\beta}(\qvec)\right)
\label{eq_comp_strain_q}
\end{equation}
in reciprocal space as already stated in Sec.~\ref{tech_fields}.
(Obviously, $\varepsilon_{\alpha\beta}=\varepsilon_{\beta\alpha}$ for any $\rvec$
in real space and any $\qvec$ in reciprocal space.)
Note that both $\epsab(\rvec)$ and its FT $\epsab(\qvec)$,
cf.~Eq.~(\ref{eq_FT_def}), are dimensionless fields.
Due to our definitions and conventions the macroscopic strain $\epsab(\qvec=\bfzero)$ 
is assumed to vanish. 
Using Eq.~(\ref{eq_comp_strain_q}) we numerically determine the three relevant components 
of the (symmetric) strain tensor field $\epsab(\qvec)$ from the two components of the displacement
field $u_{\alpha}(\qvec)$ stored on the reciprocal space grid (cf.~Fig.~\ref{fig_grid}) 
and the wavevector $q_{\alpha}$ according to Eq.~(\ref{eq_FT_gridpoints}).
As noted in Sec.~\ref{tech_fields}, in NRC there are only two
non-vanishing strain fields, namely the longitudinal and transverse strain fields
$\isflongi(\qvec)$ and $\isftrans(\qvec)$ linearly related to the 
corresponding displacement fields $\ulongi(\qvec)$ and $\utrans(\qvec)$,
cf. Eq.~(\ref{eq_tech_isflongi}) and Eq.~(\ref{eq_tech_isftrans}).

\subsection{Correlation function fields}
\label{app_comp_corr}

Using the strain tensor fields $\epsab(\qvec)$ in reciprocal space computed according to
Eq.~(\ref{eq_comp_strain_q}) for each $c$ and $t$ we obtain the correlation functions 
$\cabcd(\qvec) = \langle \epsab(\qvec) \epscd(-\qvec) \rangle$ averaged over all $c$ and $t$. 
For the reported correlation functions 
$\cabcd^{\prime}(\qvec) = \langle \epsab^{\prime}(\qvec) \epscd^{\prime}(-\qvec) \rangle$
in a coordinate system turned by an angle $\alpha$ we first compute the new 
components $u_{\alpha}^{\prime}(\rvec)$ and $q^{\prime}_{\alpha}$ of 
displacement field and wavevector. (Alternatively, one may also rotate $\epscd(\qvec)$.)
For the ICFs $\cone(q)$ and $\cfour(q)$ obtained using NRC we first
get the longitudinal and transverse displacement fields $\ulongi(\qvec)$ and $\utrans(\qvec)$ in NRC
and from those using Eq.~(\ref{eq_tech_isflongi}) and Eq.~(\ref{eq_tech_isftrans}) 
the longitudinal and transverse strains $\isflongi(\qvec)$ and $\isftrans(\qvec)$.
$\cone(q)=\la \isflongi(\qvec) \isflongi(-\qvec) \ra_{\qhatvec}$ and
$\cfour(q) = \langle \isftrans(\qvec) \isftrans(-\qvec) \rangle_{\qhatvec}$ 
are computed by averaging over all $c$ and $t$ and all wavevectors $\qvec$ with magnitude $|\qvec|$
within a chosen bin around $q$.
The correlation functions $\cabcd(\rvec)$ in real space (either for unrotated or $\alpha$-rotated 
coordinate systems) are finally obtained by inverse FFT.

\section{From $\cone(q)$ and $\cfour(q)$ to $\cabcd(\rvec)$}
\label{app_Cabcd}

As shown in Appendix~\ref{ten_field}, a forth-order tensor field describing an isotropic achiral system in two 
dimensions is given by Eq.~(\ref{eq_ten_field_o4}) in terms of four invariants $i_n(q)$. 
In turn these invariants are expressed in terms of the alternative set of 
invariants $\cone(q)$, $\cfour(q)$, $\ctwo(q)$ and $\cfour(q)$.
Due to Eq.~(\ref{eq_ICFs_vanishing}) we have $\ctwo(q) = \cthree(q) = 0$ for strain correlations.
The correlation function $\cabcd(\qvec)$ in reciprocal space are thus given by the invariants
\begin{eqnarray}
-\frac{i_1(q)}{2}=\frac{i_3(q)}{2} =i_2(q) & = & \cfour(q) \mbox{ and } \label{eq_Cabcd_i1i2i3_A} \\
i_4(q) & = & \cone(q)-4\cfour(q).  \label{eq_Cabcd_i4_A}
\end{eqnarray}
More specifically, this implies
\begin{eqnarray}
\cxxxx(\qvec) & = & c^4     \cone(q) + 4 s^2c^2        \cfour(q) \nonumber \\
\cyyyy(\qvec) & = & s^4     \cone(q) + 4 s^2c^2        \cfour(q)  \nonumber \\
\cxxyy(\qvec) & = & c^2 s^2 \cone(q) - 4 s^2c^2        \cfour(q) \nonumber \\
\cxyxy(\qvec) & = & c^2 s^2 \cone(q) + (c^2-s^2)^2     \cfour(q) \nonumber\\ 
\cxxxy(\qvec) & = & c^3 s   \cone(q) - 2 s c (c^2-s^2) \cfour(q) \nonumber\\ 
\cxyyy(\qvec) & = & c s^3   \cone(q) + 2 s c (c^2-s^2) \cfour(q) 
\label{eq_Cabcd_q}
\end{eqnarray}
with $c=\cos(\theta)=\qhat_1$ and $s=\sin(\theta)=\qhat_2$ being coefficients 
depending only on the wavevector angle $\theta$. Alternatively, 
the six relations Eq.~(\ref{eq_Cabcd_q}) may also be obtained using that the components $\epsab(\qvec)$ 
in the original coordinate frame can be expressed as
\begin{eqnarray}
\epsxx(\qvec) & = &  
c^2 \ \epsxx^{\circ}(\qvec) + s^2 \ \epsyy^{\circ}(\qvec) - 2s c \ \epsxy^{\circ}(\qvec) \nonumber \\
& = & c^2 \ \isflongi(\qvec) - 2 s c \ \isftrans(\qvec) \label{eq_eps_histo_epsxx} \\ 
\epsyy(\qvec) & = &
s^2 \ \epsxx^{\circ}(\qvec) + c^2 \ \epsyy^{\circ}(\qvec) + 2s c \ \epsxy^{\circ}(\qvec) \nonumber \\
& = &  s^2 \ \isflongi(\qvec) + 2 s c \ \isftrans(\qvec) \label{eq_eps_histo_epsyy} \\ 
\epsxy(\qvec) & = &
s c \ \epsxx^{\circ}(\qvec) - s c \ \epsyy^{\circ}(\qvec) + (c^2-s^2) \ \epsxy^{\circ}(\qvec) \nonumber
 \\
& = &  c s \ \isflongi(\qvec) + (c^2-s^2) \ \isftrans(\qvec) \label{eq_eps_histo_epsxy}
\end{eqnarray}
in terms of the longitudinal and transverse strains $\isflongi(\qvec)$ and $\isftrans(\qvec)$
and the fact that $\isflongi(\qvec)$ and $\isftrans(\qvec)$ fluctuate independently.
For the correlation functions $\cabcd^{\prime}(\qvec)$ in rotated coordinate systems 
one simply replaces $\theta$ by $x=\theta-\alpha$.
Note also that $\cxxxy^{\prime}(\qvec)$ and $\cxyyy^{\prime}(\qvec)$ do in general not vanish for all $x$ 
in standard (unrotated or rotated) coordinates. The values for NRC are obtained by setting $x=0$, 
i.e. $s=0$ and $c=1$.
We expand the angle-dependent coefficients before $\cone(q)$ and $\cfour(q)$ 
in terms of the planar harmonic functions
$\cos(p \theta)$ and $\sin(p \theta)$ with $p$ being integers
using standard trigonometric relations \cite{abramowitz}.
This implies for example
\begin{eqnarray}
\cxyxy & = & \frac{1}{8}\left[
\left(4\cfour-\cone\right) \cos(4\theta) + \left(4\cfour+\cone\right)
\right]\nonumber\\
\cxxyy & = & \frac{1}{8}\left[\left(4\cfour-\cone\right) \cos(4\theta) - \left(4\cfour-\cone\right) 
\right]\nonumber\\
\frac{\cxxxx+\cyyyy}{2} & = & \frac{1}{8}\left[
-\left(4\cfour-\cone\right) \cos(4\theta) + 3\cone+4\cfour
\right]\nonumber\\
\frac{\cxxxx-\cyyyy}{2} & = & \frac{1}{4}\left(2 \cone \right) \cos(2\theta) \nonumber
\end{eqnarray}
where we have omitted the arguments $\qvec$ on the l.h.s. and $q$ on the r.h.s.
The prefactor of $\cos(4\theta)$ and $\sin(4\theta)$ is always proportional to $i_4(q)$.
Having in mind the equipartition relation Eq.~(\ref{eq_intro_ICF_qlow}) 
and using the creep compliances $\Jone$ and $\Jtwo$ introduced in Eq.~(\ref{eq_res_JoneJtwo})
one sees that
\begin{eqnarray}
\beta V [4\cfour(q)-\cone(q)] & \to & \Jone \equiv \frac{1}{\mu}-\frac{1}{\lambda+2\mu},
\label{eq_Cabcd_Jone_def} \\
\beta V [2\cone(q)] & \to & \Jtwo \equiv \frac{2}{\lambda+2\mu} 
\label{eq_Cabcd_Jtwo_def}
\end{eqnarray}
in the low-$q$ limit. We thus get in reciprocal space
\begin{eqnarray}
\beta V \cxyxy(\qvec) & \to & \ \ \frac{\Jone}{8} \cos(4\theta) + \ldots \nonumber \\
\beta V \cxxyy(\qvec) & \to & \ \ \frac{\Jone}{8} \cos(4\theta) + \ldots \nonumber \\
\beta V \frac{\cxxxx(\qvec)+\cyyyy(\qvec)}{2} & \to & -\frac{\Jone}{8} \cos(4\theta) + \ldots \nonumber \\
\beta V \frac{\cxxxx(\qvec)-\cyyyy(\qvec)}{2} & \to & \ \ \frac{\Jtwo}{4} \cos(2\theta) \nonumber
\end{eqnarray}
where the dots mark constant terms.
These terms are irrelevant for the inverse FT, only leading to contributions at 
the origin $\rvec=\bfzero$.
\begin{figure}[t]
\centerline{\resizebox{0.9\columnwidth}{!}{\includegraphics*{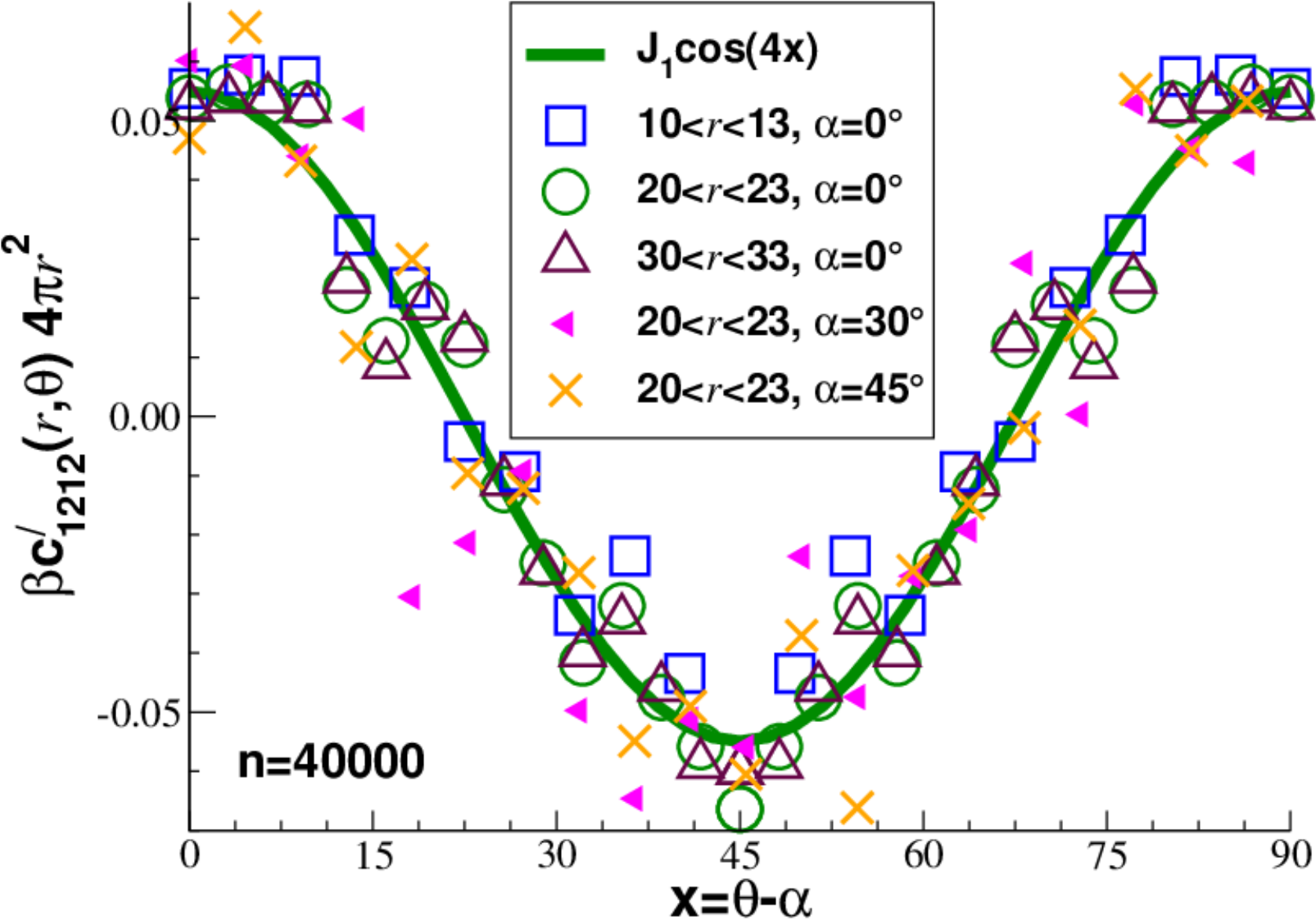}}}
\caption{Rescaled shear-strain autocorrelation function $\beta \cxyxyprime(r,\theta) 4\pi r^2$
in real space as a function of $x=\theta-\alpha$ comparing data for different $r$-intervals and 
rotation angles $\alpha$ with the prediction (bold solid line).
}
\label{fig_Cabcd_rtheta}
\end{figure}

We may thus take advantage of the analytical result for the inverse FT Eq.~(\ref{eq_q2r_fr_p}).
This leads to
\begin{eqnarray}
\beta \cxyxy(\rvec) & \to & \ \ \frac{\Jone}{4\pi r^2} \cos(4 \theta) \label{eq_Cabcd_cxyxy} \\
\beta \cxxyy(\rvec) & \to & \ \ \frac{\Jone}{4\pi r^2} \cos(4 \theta) \label{eq_Cabcd_cxxyy} \\
\beta \frac{\cxxxx(\rvec)+\cyyyy(\rvec)}{2} & \to & -\frac{\Jone}{4\pi r^2} \cos(4 \theta) 
\label{eq_Cabcd_mean} \\
\beta \frac{\cxxxx(\rvec)-\cyyyy(\rvec)}{2} & \to & -\frac{\Jtwo}{4\pi r^2} \cos(2 \theta) 
\label{eq_app_Cabcd_diff} 
\end{eqnarray}
with $\theta$ denoting the polar angle of the field vector $\rvec$.
The correlation functions $\cabcd^{\prime}(\rvec)$ in rotated coordinate systems 
generalize the above equations by substituting $\theta$ with the angle difference $x=\theta-\alpha$.
These results can be rewritten compactly using the general form expected 
from Eq.~(\ref{eq_ten4_d2_r}) for a manifest two-dimensional, isotropic and achiral 
forth-order tensor field in real space yielding 
\begin{eqnarray}
\beta \cabcd^{\prime}(\rvec) & = &
\tilde{i}_1(r) \ \ \delta_{\alpha\beta} \delta_{\gamma\delta} \label{eq_Cabcd_isomanifest} \\
& + & \tilde{i}_2(r) \ \left[
\delta_{\alpha\gamma} \delta_{\beta\delta} + \delta_{\alpha\delta} \delta_{\beta\gamma}
\right] \nonumber \\
& + & \tilde{i}_3(r) \ \left[
\rhat_{\alpha}^{\prime} \rhat_{\beta}^{\prime}\delta_{\gamma\delta} + \rhat_{\gamma}^{\prime} \rhat_{\delta}^{\prime}\delta_{\alpha\beta} 
\right] \nonumber \\
& + & \tilde{i}_4(r) \ \ \rhat_{\alpha}^{\prime} \rhat_{\beta}^{\prime} \rhat_{\gamma}^{\prime} \rhat_{\delta}^{\prime} \nonumber
\end{eqnarray}
where the invariants $\tilde{i}_n(r)$ in real space are given by
\begin{eqnarray}
4\pi r^2 \ \tilde{i}_1(r) & = & \Jtwo-3\Jone, \label{eq_Cabcd_invariants} \\
4\pi r^2 \ \tilde{i}_2(r) & = & \Jone, \nonumber \\
4\pi r^2 \ \tilde{i}_3(r) & = & 4\Jone-\Jtwo \mbox{ and } \nonumber \\
4\pi r^2 \ \tilde{i}_4(r) & = & -8\Jone. \nonumber
\end{eqnarray}
Since $i_1=-(2\Jone+\Jtwo)/4$, $i_2=(2\Jone+\Jtwo)/8$, $i_3=(2\Jone+\Jtwo)/4$ and $i_4=-\Jone$,
this is consistent with the more general relation Eq.~(\ref{eq_ten_d2_intilde}).

The angle dependence for the shear-strain autocorrelation function in real space is investigated
in Fig.~\ref{fig_Cabcd_rtheta} where we plot using linear coordinates
$\beta \cxyxyprime(r,\theta) 4\pi r^2$ as a function of $x$ for different $r$ and $\alpha$.
(To obtain sufficiently high statistics we need to average over the indicated finite $r$-bins.
This is done by weighting each data entry for a bin with the proper factor $4\pi r^2$.)
The data compare well with the prediction, Eq.~(\ref{eq_Cabcd_cxyxy}), confirming thus especially the scaling with angle difference $x=\theta-\alpha$.
Naturally, the statistics deteriorates with increasing $r$
due to the faster decay of the correlations as compared to the noise.



\end{document}